\documentclass[useAMS,usenatbib]{mn2e}
\usepackage{epsfig}

\def\d{{\rm d}}
\def\clight{{\rm c}}

\title{X-rays and hard UV radiation From the First Galaxies: \\ Ionization Bubbles and 21 cm Observations}
\author[Aparna Venkatesan \& Andrew Benson]{Aparna Venkatesan$^1$ \& Andrew Benson$^2$ \\
$^1$ Department of Physics and Astronomy, University of San Francisco, 2130 Fulton Street, San Francisco, C~ 94117, U.S.A. \\ (e-mail: avenkatesan@usfca.edu) \\
$^2$ California Institute of Technology, MC 350-17, 
1200 E. California Blvd., Pasadena, CA 91125}

\begin{document}
\maketitle
\begin{abstract}

The first stars and quasars are known sources of hard ionizing radiation in the first billion years of the Universe. We examine the joint effects of X-rays and hard UV radiation from such first-light sources on the hydrogen and helium reionization of the intergalactic medium (IGM) at early times, and the associated heating. We study the growth and evolution of individual H~II, He~II and He~III regions around early galaxies with first stars and/or QSO populations. We find that in the presence of helium-ionizing radiation,  X-rays may not dominate the ionization and thermal history of the IGM at $z \sim$ 10--20, contributing relatively modest increases to IGM ionization, and heating up to $\sim$ $10^3$--$10^5$~K in IGM temperatures. We also calculate the 21 cm signal expected from a number of scenarios with metal-free starbursts and quasars in varying combinations and masses at these redshifts. The peak values for the spin temperature reach $\sim$ 10$^4$--10$^5$ K in such cases. The maximum values for the 21 cm brightness temperature are around 30--40 mK in emission, while the net  values of the 21 cm absorption signal range from $\sim$ a few to 60 mK  on scales of 0.01--1 Mpc.  We find that the 21 cm signature of X-ray versus UV ionization could be distinct, with the emission signal expected from X-rays alone occurring at smaller scales than that from UV radiation, resulting from the inherently different spatial scales at which X-ray and UV ionization/heating manifest. This difference is time-dependent, and  becomes harder to distinguish with an increasing  X-ray contribution to the total ionizing photon  production. Such differing scale-dependent contributions from X-ray and UV photons  may therefore  ``blur" the 21 cm signature of the percolation of ionized bubbles around early halos (depending on whether a cosmic X-ray or UV background built up first), and affect the interpretation of  21 cm data constraints on reionization.

\end{abstract}

\begin{keywords}
stars: Population III.  
galaxies: high-redshift.  
(galaxies:) quasars: general.  
galaxies: star formation. 
cosmology: theory. 
(cosmology:) dark ages, reionization, first stars.  
\end{keywords}

\section{Introduction}

The first billion years after the Big Bang represents a period of great interest for studies of both galaxy formation and the evolution of the Universe as a whole. This period sees the formation of the first galaxies \citep{wise_resolving_2008} and, consequently, the beginning and completion of the process of reionizing the Universe \citep{loeb_reionization_2001,loeb_frontier_2009} as a result of the copious number of ionizing photons emitted by these sources. Current and future facilities aim to probe this epoch of the Universe both using traditional methods such as surveying faint galaxies (e.g. the James Webb Space Telescope; \citealt{gardner_james_2009}) and using novel techniques such as 21cm cosmology \citep{furlanetto06a} to probe the distribution of neutral hydrogen during the process of reionization. Understanding this epoch of the Universe from a theoretical perspective therefore requires an understanding both of the sources of ionizing photons and of the thermal and ionization state of the intergalactic medium (IGM) at these times.

Additionally, the thermal and ionization history of the IGM as a function of cosmic redshift, $z$, strongly affects the ``visibility'' of the most distant galaxies and quasars \citep{madau_radiative_1995,meiksin_colour_2006,dayal_visibility_2011}, and the feedback exerted on the formation of new galaxies \citep{efstathiou_suppressing_1992,quinn_photoionization_1996,navarro_effects_1997,barkana_photoevaporation_1999,bullock_reionization_2000,somerville_can_2002,
Benson:02a,Benson:02b,koposov_quantitative_2009,munoz_probing_2009,busha_impact_2010,maccio_luminosity_2010}.  The process of reionization is expected to begin with the formation of ionized bubbles around luminous sources in the redshift range $z=10$--20. These bubbles will eventually grow in size and number until complete overlap is reached and the Universe becomes fully reionized. The shapes and sizes of bubbles will be controlled by the cosmological density field and the process of galaxy formation. Their internal ionization and temperature structure will depend on the spectrum of the input source (i.e. how hard the photons are) and the efficiencies of recombination and cooling processes.

The recent data from WMAP-7 \citep{Larson:10} reveal that the IGM is fully ionized up to $ z \sim 10$, most likely  with a  period of partial ionization at higher redshifts.  
Theoretical work over the last fifteen years has focused mostly on the hydrogen reionization of the IGM \citep{gnedin_reionization_1997,chiu_semianalytic_2000,ciardi_inhomogeneous_2000,somerville_epoch_2003,onken_history_2004,benson_epoch_2006,furlanetto06a}. However, helium reionization has received comparatively less attention, ranging from calculations of helium/hydrogen reionization from the first stars and QSOs at $z \ga$ 6 \citep{Venkatesan:03a,Wyithe:03} to studies of helium reionization by QSOs at $ z \sim$ 3 \citep{sokasian03,furlanetto08b,furlanetto08a}. Although helium is the second most abundant element, its substantially higher ionization energy relative to hydrogen, as well as its interactions with X-rays through secondary ionizations, can lead to significant effects for the high-$z$ IGM and the cosmic microwave background (CMB) once reionization has occurred even to a partial degree.  Additionally,
X-rays have greater penetrating power relative to UV radiation. When occurring in combination with helium ionization from the first stars and quasars, X-rays could act to strongly alter the ionization and thermal history of the IGM. 

 In this work we investigate the joint impact of X-rays and helium-ionizing radiation from the first galaxies on IGM reionization and  heating. We focus on the growth and evolution of individual ionization fronts in H and He, rather than a fully evolving cosmological calculation, which we plan to pursue in future work (\S 4). We study whether  the differing contributions arising from X-rays versus UV ionization can be distinguished through 21 cm observations.  
Recent papers by other authors have focused on specific aspects of this problem in other contexts, e.g., helium reionization by quasars at lower redshifts ($z \sim 3$) \citep{Bolton:09, McQuinn:09, furlanetto08a}, without explicitly considering the effects of X-ray heating \citep{furlanetto08b}, or, with only a single high-mass star embedded
in a high-z galaxy halo \citep{chen08}. We will demonstrate that X-rays \emph{may not} play a dominant role in high-$z$ ionization, contrary to the theoretical expectations in some previous works (see, e.g., \citealt{Thomas:08}), and will ask the question: does there exist a cosmological epoch when the IGM's thermodynamic and ionization properties are determined mostly by X-rays? 

The tradeoff of these ionization effects will have important consequences for predictions for future radio observations that plan to see ionized bubbles in emission or absorption against the CMB. There has already been a substantial body of work on the feedback on ionization \citep{Venkatesan:03a,tumlinson03,Wyithe:03} and emission line signatures \citep{Oh:01,tumlinson01,Venkatesan:03a,Dawson:07} arising from first-light sources that have hard ionizing spectra. Here, we focus on the radio signatures as  the topology of reionization arising from X-rays versus UV radiation is expected to be different. We also test other theoretical predictions for the growth of individual ionized regions around early galaxies, e.g., that for sufficiently hard sources such as the first stars and QSOs, the H and He I-fronts may track each other closely.

The remainder of this paper is arranged as follows. In Section 2, we describe the model that we use to follow the growth of cosmological ionization fronts around evolving sources. In Section 3, we present our results for the thermal and ionization properties of such regions,  their observable signatures (including 21 cm signals) around a set of representative sources, and compare our findings with earlier works in this field. We conclude in Section 4.

\section{Background and Models}

We assume a background cosmology using the most recent cosmological parameters fits
from the WMAP-7 CMB data \citep{Larson:10}. We combine the formalism for studying the
non-equilibrium evolution of hydrogen and helium in the IGM in \citet{Venkatesan:03a} and \citet{tumlinson04}, and the
input processes related to X-ray ionization in the high-z IGM in \citet{Venkatesan:01}  with the
code Galacticus. Galacticus is a newly developed semi-analytic code on galaxy formation \citep{Benson:11} which includes feedback from high-redshift 
star/quasar formation while meeting current experimental constraints at lower redshifts.  Here, we have utilized it to solve for the growth of a spherical ionization front around a point
source in the IGM. The ionizing and heating processes included in 
this code are described in detail below. 

We are primarily interested in the effects of hard ionizing radiation from the first galaxies - these are assumed to be of order $10^8$ M$_\odot$ in total mass and of approximate size 1 to a few kpc\footnote{A 10$^{10}$  M$_\odot$ halo at $z=10$ has an approximate physical (not comoving) virial radius of $\sim$ 7 kpc, with a galaxy of size $\sim$ 1 kpc in it. A massive Milky Way-mass halo  (10$^{12}$ M$_\odot$) at that redshift would be about 33 kpc, with a typical galaxy of a few kpc in size.}. We follow the advancing ionized fractions\footnote{This is  in contrast to \citet{Venkatesan:01}, where the average IGM ionization fraction was computed without tracking the growth of individual I-fronts around the halos containing the QSO.} around a starburst and/or quasar in such a halo, and treat the IGM as being homogenous around the source. In particular, we do not include a density enhancement as would be expected if the source forms in
the center of a dark matter halo. In general, the ionization fronts we find
are much larger than the sizes of typical halos at these redshifts and so will
be insensitive to the details of the density profile on small scales.
Additionally, sources such as those considered here will likely form in halos
sufficiently massive to collisionally ionize hydrogen and helium, such that
the photoionization front would begin growing from the edge of the
collisionally ionized region\footnote{In a fully 3-D calculation these halos
would accrete most of their mass via cold filaments of gas which are not shock
heated as they enter the halo and so are not collisionally ionized. It is
beyond the scope of this work to examine the effect of such filaments on the
growth of ionization fronts \citep{Keres05}, but they can be expected to impede the
growth of the front along directions coinciding with a filament, while
permitting faster growth along directions between filaments.}.

We consider quasars with varying black hole (BH) masses, and model a typical QSO spectrum with the fit given in Haardt \& Madau (1996).  We assume that the duty cycle of the QSO is 100 million years typically - significantly longer duty cycles would exceed the Hubble time at $z \sim$ 10--20.  In our models, we allow the AGN to be on for 100 Myr before it is shut  off. We include the effects of metal-free stars occurring in starbursts of varying masses --- the fits are taken from \citet{Venkatesan:03a}.  

The non-equilibrium ionization fractions are calculated including the following processes: photoionization, collisional ionization, case B radiative recombination, dielectronic recombination for He~I, and the coupling between H and He caused by the  radiation fields from the He~I 24.6 eV recombination continuum and from the bound-bound transitions of He~I \citep{Venkatesan:01}.  The photoionization cross sections for H I and and He II are taken from Spitzer (1978), and from Verner et al. (1996) for He~I. The ratio of the H~I to He~I photoionization cross sections decreases with photon energy, ranging from about 5\% at 100 eV to 3.5\% at 1 keV. This implies that an X-ray photon is ``seen" better by a He~I atom than by a H~I atom.

We also include secondary ionizations and excitations of H I and He I arising from the X-rays (Shull \& van Steenberg 1985). As noted in \citet{Venkatesan:01}, a typical X-ray photon is far more likely to be absorbed by He I rather than H I, so that secondary ionization (rather than direct photoionization) is most relevant for H~I when X-rays dominate  photoionization. The resulting photoelectrons will ionize many more H~I atoms than He~I, H~I atoms being more numerous. As the background ionization increases, the photoelectron deposits more and more of its energy in heat and less in collisional ionizations/excitations. Shull \& van Steenberg (1985) assumed that  the ionization fractions of H~I and He~I were equal, and we have replaced the generic ionization fraction in their formulae with the electron fraction $x_e$ which is more directly relevant for the IGM. 

The thermal evolution of the gas is computed including the following processes \citep{Venkatesan:01}: photoelectric heating from the secondary electrons of H and He, which is itself a function of the background ionization levels (Shull \& van Steenberg 1985), and, heating from the H I photoelectrons liberated by the bound-bound transitions or the 24.6 eV recombination continuum of He~I. Cooling terms include radiative and dielectronic recombination (\citealt{Venkatesan:01}  and references therein), thermal bremsstrahlung, Compton scattering off the CMB, collisional ionization and excitation, and the adiabatic expansion of the IGM.  The contributions to heating and cooling from the scattering of the secondary Ly$\alpha$ photons from X-ray ionization is negligible \citep{chen04,chen08} and is not included here.

Our 1D non-equilibrium ionization code includes all of the above ionization and heating processes, and solves for the evolution of the thermal and ionization state around the source as follows. The IGM surrounding the source is divided up into a large number of concentric spherical shells. Unless otherwise noted, we use 1000 shells, spaced logarithmically in radius from $10^{-4}$ to $10$~Mpc. These shells are initially populated with hydrogen and helium in a primordial ratio.

When considering a uniform medium surrounding the source, the gas is given initial ionized fractions as determined by the {\sc RecFast} recombination code\footnote{We use v1.4.2 of {\sc RecFast} and include all of the modifications to the HeI recombination rate.} \citep{seager_how_2000} for the appropriate cosmology and redshift. The initial temperature of the gas in each shell is also determined by {\sc RecFast} and each shell is initially set to be expanding with the Hubble flow.

We then proceed to evolve the thermal and ionization states of these shells forwards in time in a series of short time steps. During each time step we begin by computing the input spectrum of photons emitted by the central source (QSO, stars or both). Given this spectrum, we compute rates of ionization and heating in the innermost shell and solve for the evolution of its properties by integrating the appropriate set of differential equations as desribed below. The input spectrum is then attenuated by the optical depth of this first shell and used as input for the second shell. This process is repeated until the outermost shell is reached (which is chosen to be at sufficiently large radius that the radiation field is attenuated to close to zero at all times during our calculation). In addition to changes in temperature and ionization state, the density of each shell evolves as it expands or contracts due to any initial velocity and pressure forces. This approach is similar to those in other recent papers, e.g., Thomas \& Zaroubi (2008).

Our calculations of the ionization and thermal evolution of each shell use the same input physics as the IGM evolution model of \cite{benson_galaxy_2010}. The density of each ionization, $n_i$, state in a given shell is then given by
\begin{eqnarray}
{\d n_i \over \d t} & = & -n_i {\dot{V}\over V} + [\alpha_i(T)n_{i+1}n_e-\alpha_{i-1}(T)n_in_e\nonumber \\
 & & -\Gamma_{e,i}(T)n_in_e +\Gamma_{e,i-1}(T)n_{i-1}n_e-\Gamma_{\gamma,i}n_i\nonumber \\
 & &+\Gamma_{\gamma,{i-1}}n_{i-1}]
\end{eqnarray}
where for each atomic species H or He, i refers to their ionization state (i.e., i = 1 and 2 for H and H$^+$, and i = 3, 4 and 5 for He, He$^+$ and He$^{2+}$), $n_i$ is the number density, $T$ is the temperature of the shell, $V$ is the volume of the shell, $\alpha_i$ is the recombination rate for $i$ \citep{verner_atomic_1996}, $\Gamma_{e,i}$ is the collisional ionization rate coefficient for $i$ \citep{voronov_practical_1997} and $\Gamma_{\gamma,i}$ is the photo-ionization rate for $i$ which is given by
\begin{equation}
 \Gamma_{\gamma,i} = \int^{\infty}_0 \sigma_i^\prime(E) n_i {S_{\gamma}(E) {\rm e}^{-\tau(E;r)} \over 4 \pi r^2} \d E,
\end{equation}
where $\sigma_i^\prime$ is an effective photo-ionization cross-section that accounts for the effects of secondary ionizations and is given by \citet{shull85} (as re-expressed by \citealt{Venkatesan:01}):
\begin{eqnarray}
 \sigma_{\rm H}^\prime(E) &=& \left(1+\phi_{\hbox{\scriptsize H{\sc i}}} { E-E_{\rm H} \over E_{\rm H}} + \phi^*_{\hbox{\scriptsize He{\sc i}}} {E-E_{\rm H}\over 19.95\hbox{eV}}\right) \sigma_{\rm H}(E) \nonumber \\
 & & + \left(1+\phi_{\hbox{\scriptsize He{\sc i}}} {E-E_{\rm He}\over E_{\rm He}}\right) \sigma_{\rm He}(E), \\
 \sigma_{\rm He}^\prime(E) &=& \left(1+\phi_{\hbox{\scriptsize He{\sc i}}} {E-E_{\rm He} \over E_{\rm He}}\right) \sigma_{\rm He}(E) \nonumber \\
 & & + \left(\phi_{\hbox{\scriptsize He{\sc i}}} {E-E_{\rm H}\over24.6}\right) \sigma_{\rm H}(E),
\end{eqnarray}
where $\sigma(E)$ is the actual cross section \citep{verner_analytic_1995} and
\begin{eqnarray}
 \phi_{\hbox{\scriptsize H{\sc i}}}   &=& 0.3908 (1-x_{\rm e}^{0.4092})^{1.7592}, \\
 \phi^*_{\hbox{\scriptsize He{\sc i}}} &=& 0.0246 (1-x_{\rm e}^{0.4049})^{1.6594}, \\
 \phi_{\hbox{\scriptsize He{\sc i}}}   &=& 0.0554 (1-x_{\rm e}^{0.4614})^{1.6660}.
\end{eqnarray}
In the above, $S(E) \d E$ is the number of photons emitted per second in the energy range $E$ to $E+\d E$ by the central source and $\tau(E;r)$ is the optical depth to radius $r$ at energy $E$.

Similarly, the evolution of the temperature of each shell is given by
\begin{equation}
{\d T \over \d t} = -(\gamma-1) T {\dot{V}\over V} + {T \over \mu}{\d \mu \over \d t} + {\left(\Sigma^T - \Lambda^T\right) \over {3 \over 2}k_{\rm B}n_{\rm tot}}.
\end{equation}
Here, $\gamma$ is the adiabatic index of the gas, $\Sigma^T$ is the rate of heating per unit volume due to all the heat sources (i.e. Compton heating and photo-heating) and $\Lambda^T$ is the rate of cooling per unit volume due to all the heat sinks (i.e. Bremsstrahlung cooling and various atomic processes), $n_{\rm tot}$ is the total number density of atoms (H and He) and their ions per unit volume, $T$ is the temperature of the shell and $k_{\rm B}$ is Boltzmann's constant.

In the above equation the first term represents adiabatic cooling due to the expansion of the shell. The second term accounts for the effects of changes in the mean atomic mass due to ionization and recombination processes. The final term accounts for the heating and cooling effects of the various processes that we now discuss below.\\

\noindent {\bf Photoheating}

Photoionization heats the shell at a rate of 
\begin{equation}
\Sigma_{\rm photo} = \int^{\infty}_0(E-E_i) \sigma^\prime(E)n_i {S_{\gamma}(E) {\rm e}^{-\tau(E;r)} \over 4 \pi r^2} {\mathcal E} \d E
\end{equation}
where $E_i$ is the energy of the sampled photons which is associated with atom/ion number density $n_i$, $\sigma^\prime$ is the effective partial photo-ionization cross section (accounting for secondary ionizations) for the ionization stages of H and He, $n_{\gamma(E)}$ is the number density of photons of energy $E$, and $E_i$ is the ionization potential of $i$. In the above, ${\mathcal E}$ accounts for heating by secondary electrons and is given by \citep{shull85}:
\begin{equation}
 {\mathcal E} = 0.9971 [1-(1-x_{\rm e}^{0.2663})^{1.3163}].
\end{equation}

\noindent {\bf Compton Cooling/Heating}

Compton scattering of CMB photons from free electrons causes cooling or heating of the gas at a rate of \citep{peebles_recombination_1968}
\begin{equation}
\Sigma_{\rm Compton} = 4\sigma_{\rm T} a_{\rm R}\left(T_{\rm CMB}(1+z)\right)^4{n_{\rm e} {\rm k}_{\rm B} \over {\rm m}_{\rm e}\clight} \left(T_{\rm CMB}(1+z) - T\right),
\end{equation}
where $\sigma_{\rm T}$ is the Thompson cross section, a$_{\rm R}$ is the radiation constant, T$_{\rm CMB}$ is the temperature of the CMB at $z=0$, $n_{\rm e}$ is the number density of electrons per unit volume and ${\rm m}_{\rm e}$ is the mass of an electron. 

For a typical source in our paper, we find that Compton heating is insignificant. The initial emission rate of ionizing photons for a $10^5$ M$_\odot$ starburst with a $10^6$ M$_\odot$ BH (detailed in the next section) is $\sim 1.3 \times 10^{51}$ photons s$^{-1}$. The radius to which Compton heating is important \citep{Ricotti:08} for this scenario at $z=10$ is about 99 pc. As we will see, this is well below the 0.001--1 Mpc scales that are most relevant for I-front evolution and 21 cm signals in this work (\S 3); thus, Compton heating will not have a significant effect on our results.

\noindent {\bf Single Electron Recombination Cooling}

Photon emission due to single electron recombination cools the shell at a rate
\begin{eqnarray}
\Lambda_{\rm rec} &=& {3 \over 4}k_{\rm B}T \left[\alpha_{\rm r_{\rm H^+}}(T)n_{\rm H^+}+\alpha_{\rm r_{\rm He^+}}(T)n_{\rm He^+} \right. \nonumber \\
 & & \left.+\alpha_{\rm r_{\rm He^{2+}}}(T)n_{\rm He^{2+}}\right] n_{\rm e},
\end{eqnarray}
where $\alpha_{\rm r}$ is the rate of the recombination processes for its respective atom/ion number densities, $n_i$ \citep{verner_atomic_1996}.\\

\noindent {\bf Dielectric Recombination Cooling}

Photon emission due to dielectric recombination cools the shell at a rate
\begin{equation}
\Lambda_{\rm dielec} = 40.74 \, \hbox{eV} \; \alpha_{\rm d}(T)n_{\rm He^{2+}}n_{\rm e}
\end{equation}
where $\alpha_{\rm d}$ is the rate of the recombination process for He$^{2+}$ \citep{aldrovandi_radiative_1973,shull_ionization_1982,arnaud_updated_1985}.\\

\noindent {\bf Collisional Ionization Cooling}

Collisional ionization leads to a cooling rate of
\begin{eqnarray}
\Lambda_{\rm ion} &=& \left[E_{\rm H}\alpha_{\rm i_{\rm H}}(T)n_{\rm H}+ E_{\rm He}\alpha_{\rm i_{\rm He}}(T)n_{\rm He} \right. \nonumber \\
 & & \left. + E_{\rm He^+}\alpha_{\rm i_{\rm He^+}}(T)n_{\rm He^+}\right]n_{\rm e},
\end{eqnarray}
where $\alpha_{\rm i}$ is the collisional ionization rate coefficient for the respective atom/ion of number density $n_i$ and $E_i$ is the ionization potential of the respective atom/ion, H, He and He$^+$.\\

\noindent {\bf Collisional Excitation Cooling}

Collisional excitation followed by radiative decay cools the shell at a rate:
\begin{equation}
\Lambda_{\rm ex} = \left(\alpha_{\rm coll_H}n_{\rm H}+\alpha_{\rm coll_{He^+}}n_{\rm He^+}\right)n_{\rm e},
\end{equation}
where $\alpha_{\rm coll H}$ and $\alpha_{\rm coll {He^+}}$ are the rates of collisional excitations involving H and He$^+$ respectively \citep{scholz_collisional_1991}.\\

\noindent {\bf Bremsstrahlung Cooling}

Finally, Bremsstrahlung emission cools the shell at a rate
\begin{eqnarray}
\Lambda_{\rm Brem} &=& {16 \over 3\sqrt3}\left({2\pi k_{\rm B} \over \hbar^2m_{\rm e}^3}\right)^{1\over2}\left({e^2 \over 4\clight\pi\epsilon_0}\right)^3\clight^2\sqrt{T}\left[\gamma_{\rm H^+}(T) n_{\rm H^+} \right. \nonumber \\
 & & \left. +\gamma_{\rm He^+}(T) n_{\rm He^+}+4\gamma_{\rm He^{2+}}(T) n_{\rm He^{2+}}\right]n_{\rm e}.
\end{eqnarray}
Here, $\epsilon_0$ is the permittivity of free space and $\gamma$ is the energy-averaged Gaunt factor \citep{sutherland_accurate_1998}.

These coupled differential equations are solved numerically using a standard Runge-Kutta method.

\section{Results}

As noted earlier, we focus on early galaxies of typical mass  $\sim 10^8$--$10^{10}$ M$_\odot$ in total mass and of approximate size a few kpc at most. We therefore perform most of our calculations at $z = 10$, with one calculation at $z = 20$ for comparison.

\begin{figure}
\includegraphics[width=3.0in]{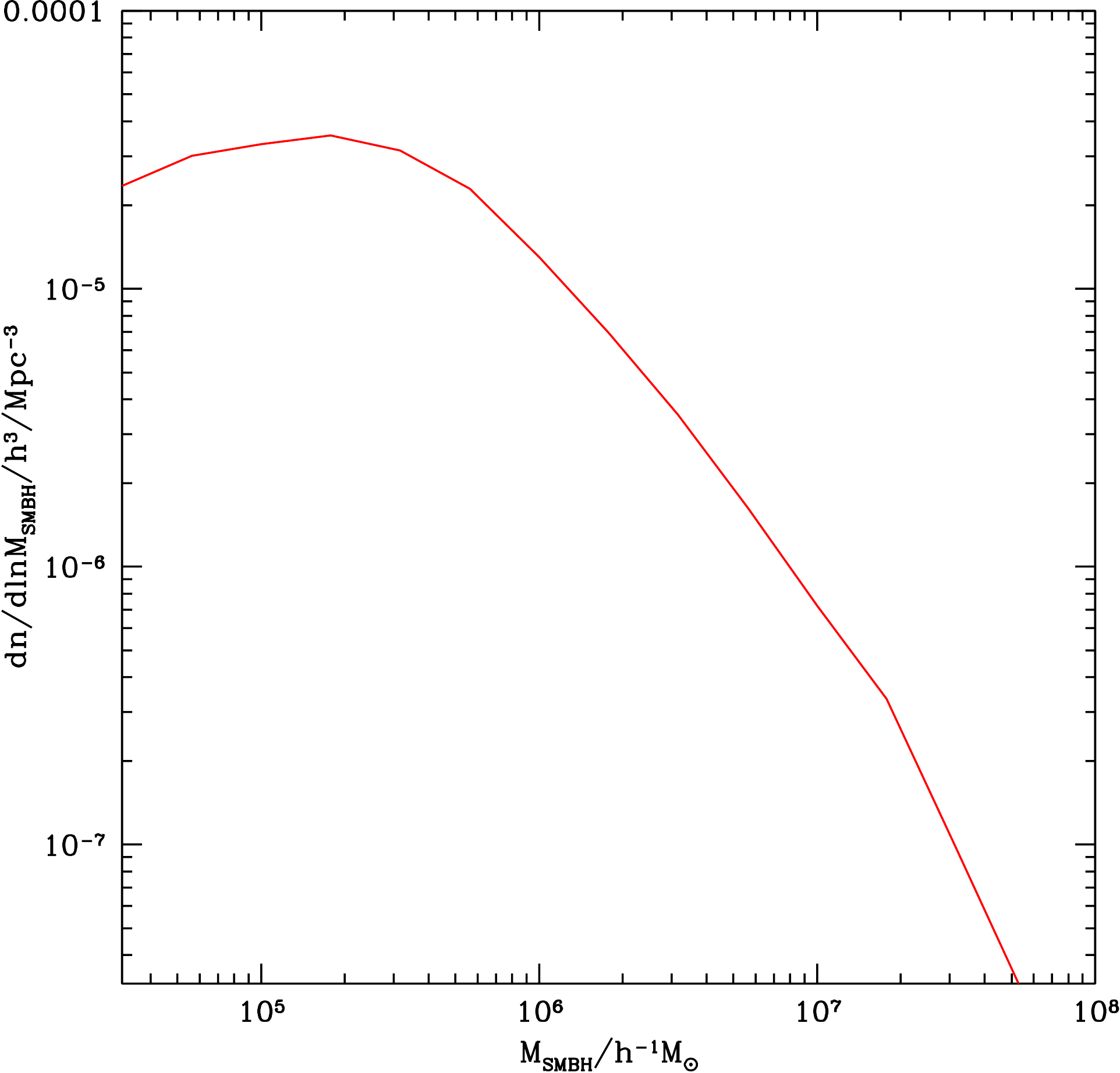}
\caption{The BH mass function at $z = 10$, using public data from the Millennium Simulation database. Note the peak around 10$^5$--10$^6$ M$_\odot$. See text for more discussion.}
\label{fig:smbh}
\end{figure}

To calculate the feedback from a typical QSO/starforming galaxy at these epochs, we compute the BH mass function at $z = 10$  using data that is publicly available from the
Millennium Simulation database\footnote{The Virgo-Millennium database is available at: http://www.g-vo.org/Millennium/} \citep{springel05}.   
In Figure ~\ref{fig:smbh}, we show the computed BH mass function at $z=10$, where we see that a typical quasar is powered by BHs in the mass range $\sim$ 10$^5$--10$^6$ M$_\odot$, which we use as a baseline for most of the cases considered in this paper. The turnover in Figure ~\ref{fig:smbh} may be partially due to the finite resolution of the simulation itself; in reality, we expect that the mass function should continue to slowly rise to somewhat smaller masses. In our models, the X-rays from the stellar populations are minimal, so we consider cases where the BH mass is typically $10^6$ M$_\odot$, with some lower BH-mass cases (down to no BH) and one case with a BH mass of $10^8$ M$_\odot$ to  derive an upper limit to the X-ray feedback. We assume that the duty cycle of the QSO is 100 Myr for nearly all our cases but include one case with a low-mass BH QSO that has a shorter duty cycle of 10 Myr.

Note that the the typical ratio of BH to stellar  burst masses considered here are not consistent with the measured ratio of the BH to stellar spheroid (bulge) mass of 0.15\% at $z=0$ \citep{gultekin09}. 
Early galaxies differ from present day ones in that they must have a seed BH that grows with time over generations of starbursts and galaxy mergers. Today we measure the BH to star (or spheroid) mass ratio {\it after} these processes have happened but it is unclear what this ratio would be for primordial galaxies, or if this ratio remains constant down to lower galaxy masses \citep{Greene10}. AGN observations indicate a possible lag in the peak of BH growth (and therefore AGN activity) relative to the peak in the star formation rate in early galaxies, owing to gas dynamical effects between star formation and BH ``feeding" \citep{Hopkins11}.  There are additional uncertainties related to the gas fraction, the Eddington ratio etc. at high redshifts. Thus, we provide a few example cases here but do not attempt to provide a cosmological sample of model galaxies.

\begin{figure}
\includegraphics[width=3.0in]{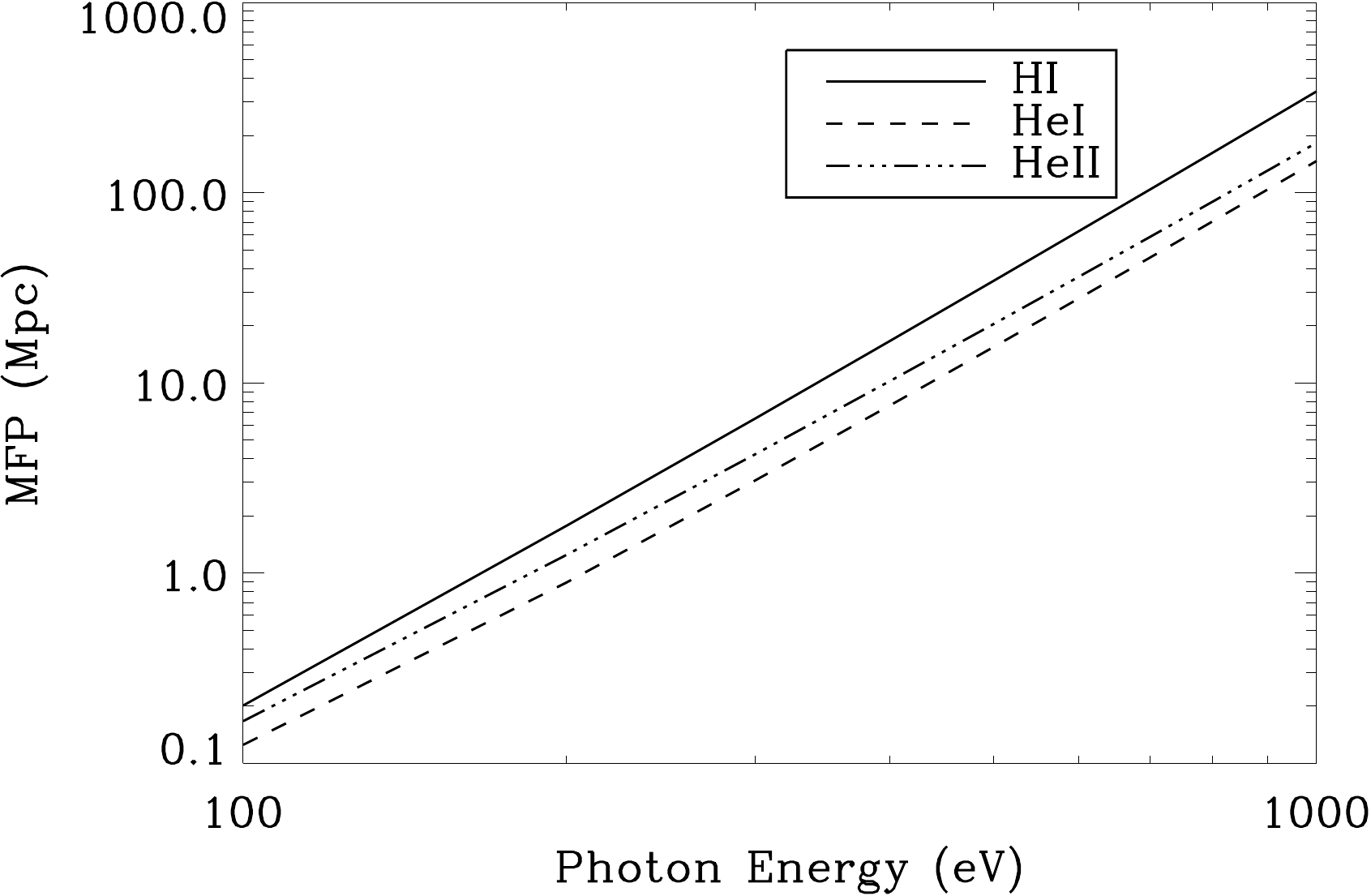}
\caption{The mean free path in Mpc for H~I, He~I, and He~II at $z=10$ for photon energies ranging from 0.1 to 1 keV.}
\label{fig:mfp}
\end{figure}

In order to distinguish the contributions of X-ray ionization relative to that from UV radiation, we consider three variations on each case with a starburst and QSO: one with the full spectrum including UV and X-ray photons from the source, one without the X-rays, and one with the X-rays alone. To do this, we need to define the boundary between what is considered an X-ray versus a hard UV photon, a quantity that has often not been clearly defined in the cosmology literature on this topic \citep{chen08,ricotti05}. At least some of this difference arises from considering the spectrum at the source versus the emergent spectrum after processing through the gas in the galaxy. We choose 120 eV as the minimum threshold for what we consider an X-ray. This is consistent with the broader physics definition, but also with the impact of a typical X-ray on the IGM. We discuss this in detail in Section 3.3, but we note for now the well-known result that the mean free path (MFP) of  X-rays varies substantially by X-ray energy. We show this explicitly in Figure ~\ref{fig:mfp}:  a 100 eV photon has a MFP of 0.1--0.2 Mpc whereas a 1 keV photon has a MFP that is larger by more than 3 orders of magnitude. Note too the ``ranking" of the three species in this plot - He~I has the lowest MFP at all energies, representing the bottleneck for X-rays that results in secondary ionizations for H~I (Section 2).

\subsection{Feedback from First Stars and QSOs}

\begin{figure}
\begin{center} $
\begin{array}{cc} 
\includegraphics[width=1.5in]{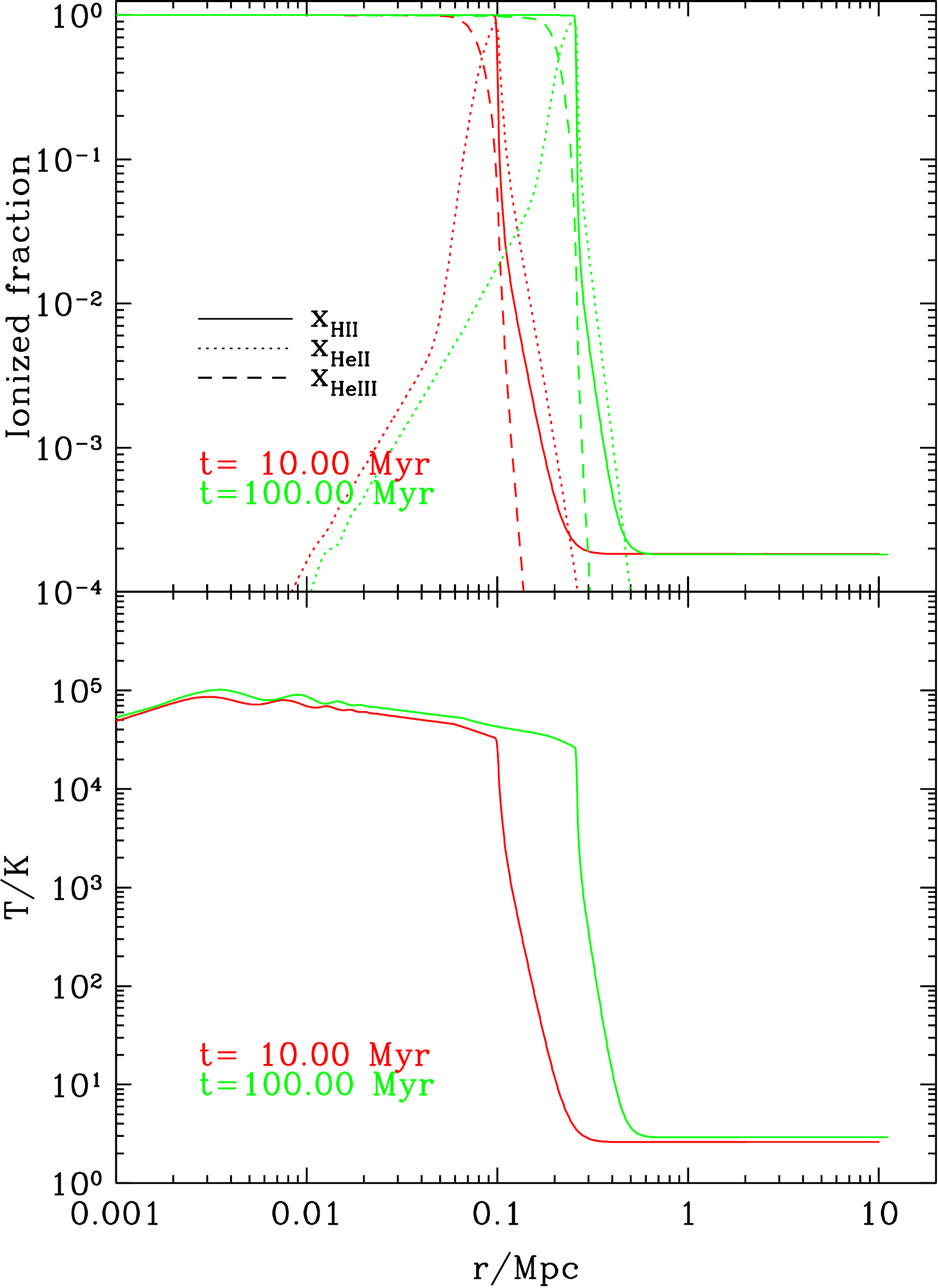} &
\includegraphics[width=1.5in]{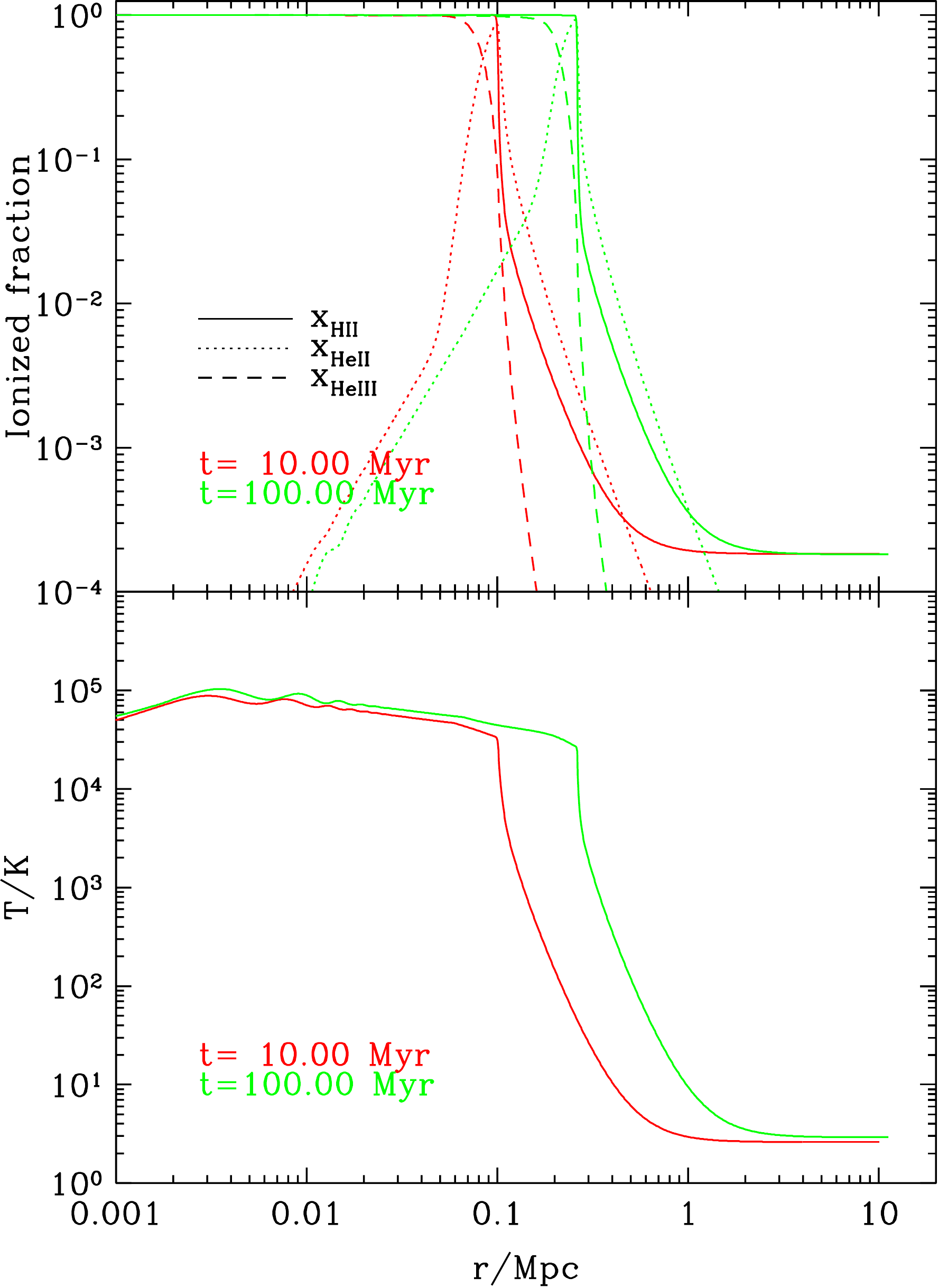} \\
\includegraphics[width=1.5in]{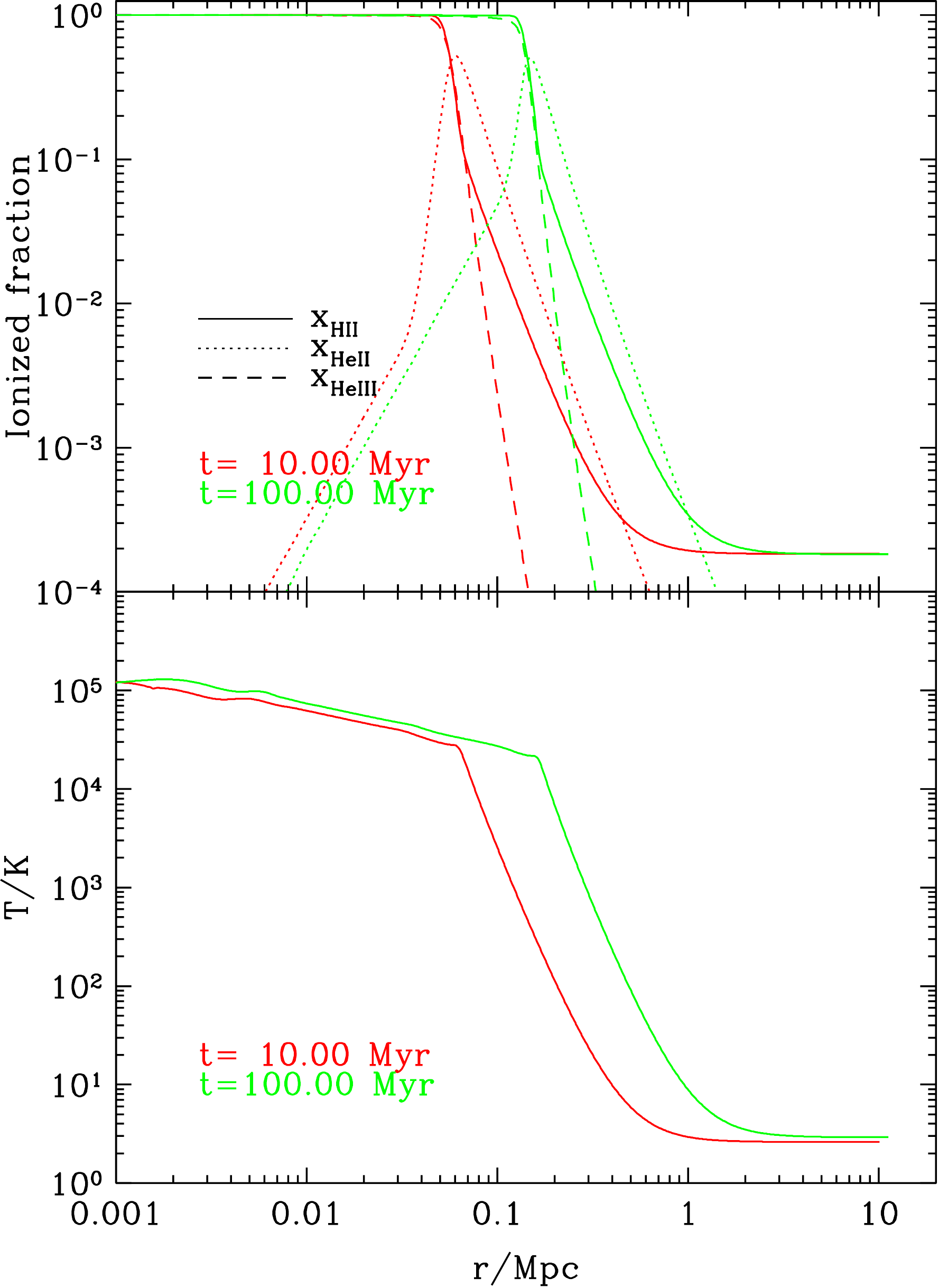} & \\
\end{array} $
\end{center}
\caption{The ionization and temperature profiles for a $10^5$ solar-mass starburst with $10^6$ solar-mass QSO/BH at  $z \sim 10$ (our standard case). The solid, dotted and dashed lines represent the fraction of H~II, He~II and He~III respectively. Red and green curves show the curves at times 10 Myr and 100 Myr after the quasar turns on. The upper left and upper right panels display the cases with the full QSO spectrum with UV photons, but that exclude and include X-rays from the central QSO. The lower panel shows the effects arising from X-rays alone.}
\label{fig:ionstd}
\end{figure}

We begin by examining a number of cases at $z = 10$ that involve varying combinations of starburst and BH masses. The plots all show cases with and without X-rays, and one with X-rays only (i.e. no lower energy photons). We begin with a $10^5$ M$_\odot$ starburst with a $10^6$ M$_\odot$ BH, hereafter referred to as the standard case. Figure ~\ref{fig:ionstd} displays the ionization and temperature profiles as a function of distance from the central starburst/QSO source at $z=10$, for the species H~II, He~II, and He~III. The red and green curves respectively show the evolution of the ionization and temperature curves at times 10 Myr and 100 Myr after the source turns on.
The X-rays contribute from $\sim$ a few percent up to full ionization in different H/He species at  IGM scales (10--100 kpc), and heating of the order $10^4$--$10^5$ K.  Although the panels with and without X-rays (the upper two panels) look very similar at first glance, we note the extended tail of low-level ionization in H~II and He~II (but not He~III) beyond the I-front: the signature of X-ray ionization. This can be seen in the red curves (10 Myr) on physical scales of 0.1--0.2 Mpc.

\begin{figure}
\begin{center} $
\begin{array}{cc} 
\includegraphics[width=1.5in]{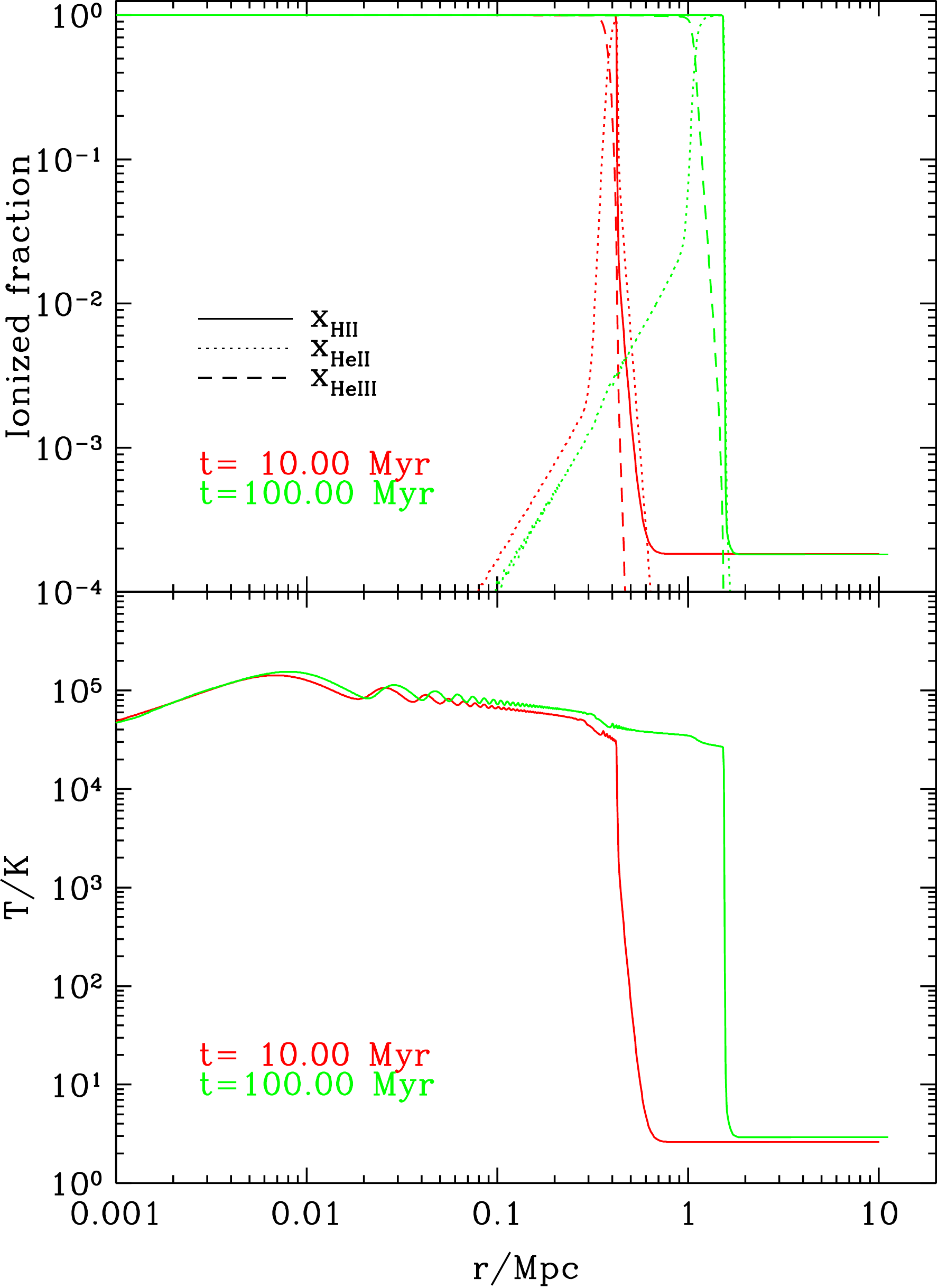} &
\includegraphics[width=1.5in]{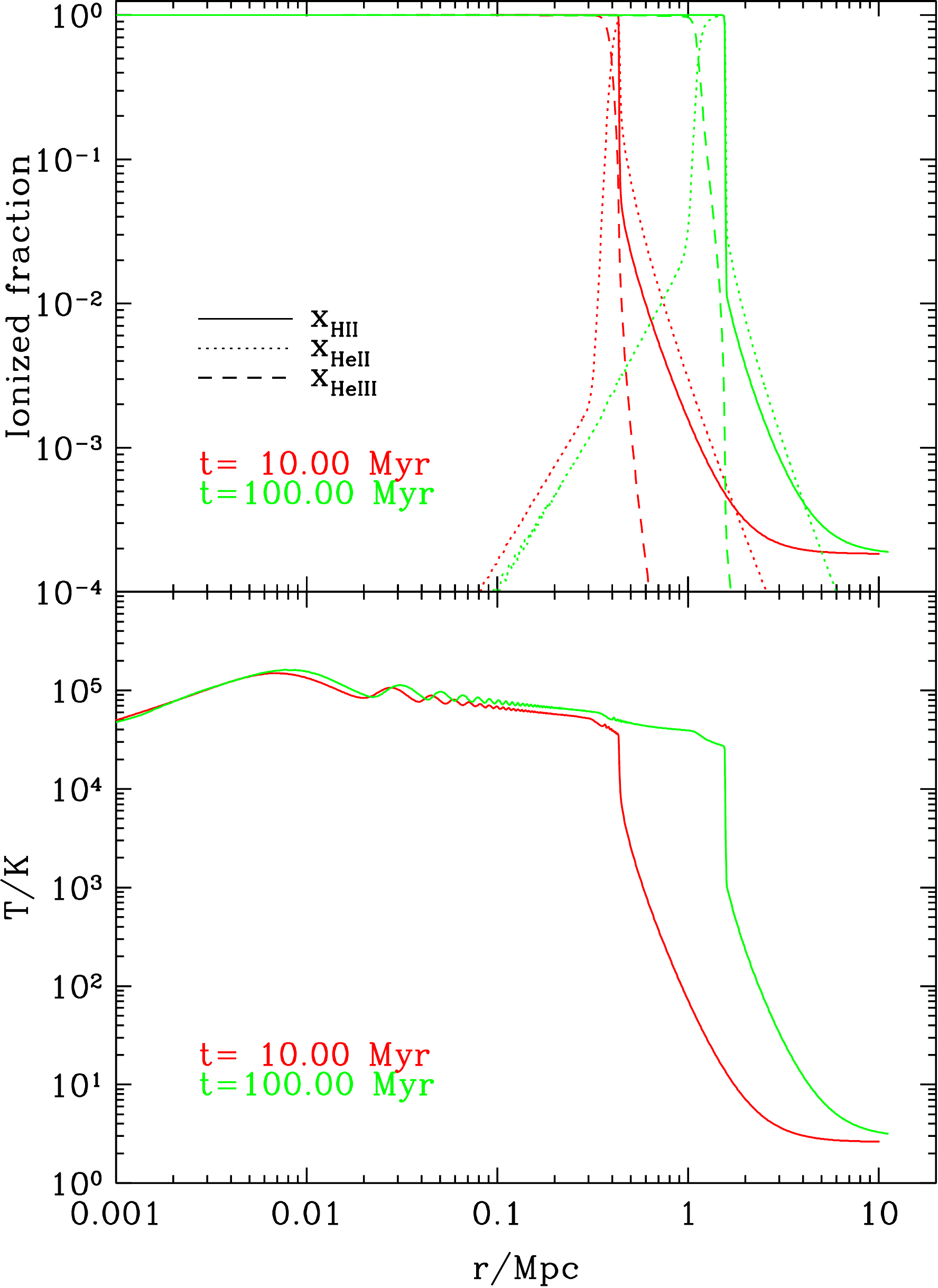} \\
\includegraphics[width=1.5in]{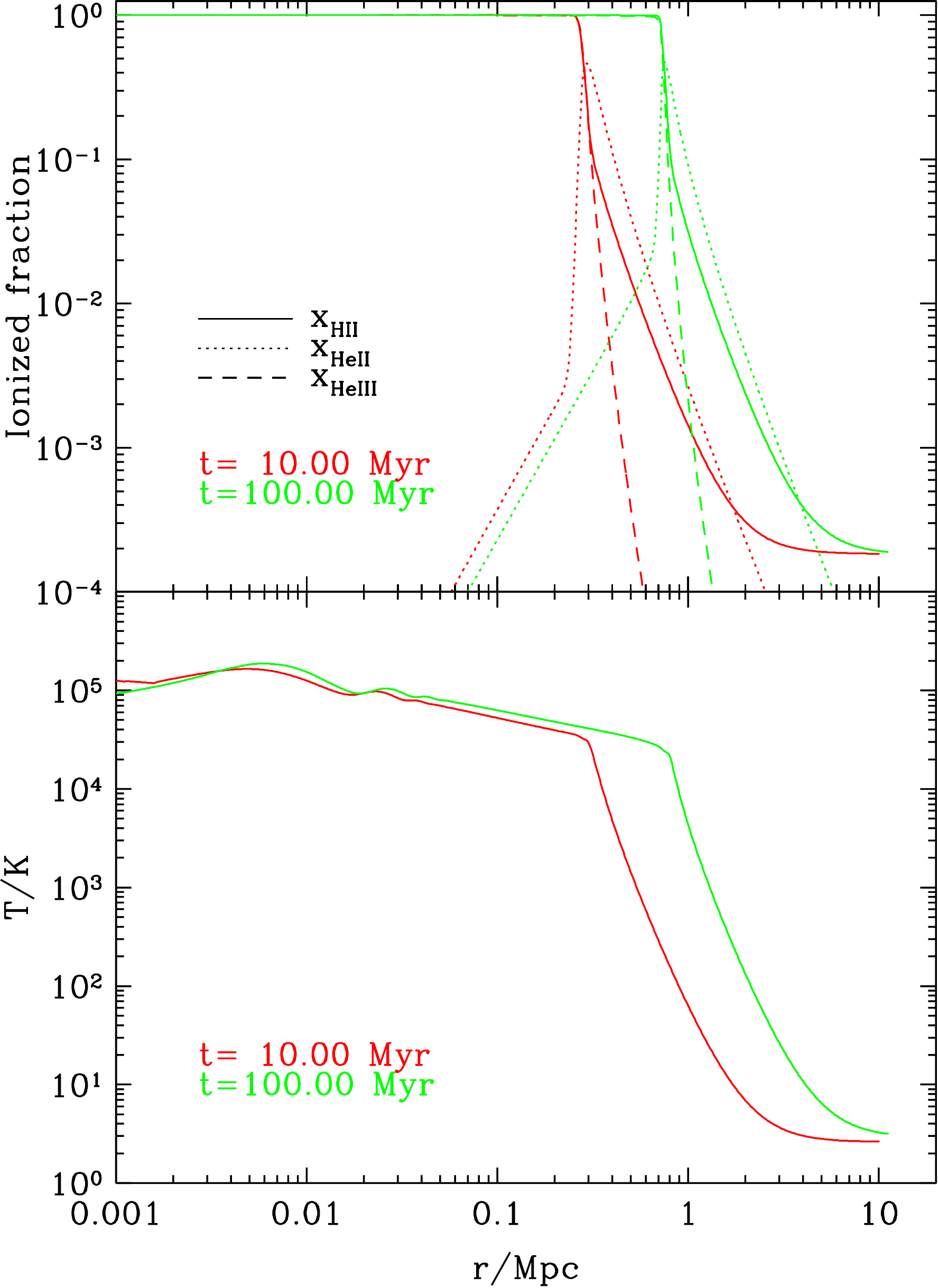} & \\
\end{array} $
\end{center}
\caption{The ionization and temperature profiles for a $10^5$ solar-mass starburst with $10^8$ solar-mass QSO/BH at  $z \sim 10$.}
 \label{fig:bigbh}
\end{figure}

\begin{figure}
\begin{center} $
\begin{array}{cc} 
\includegraphics[width=1.5in]{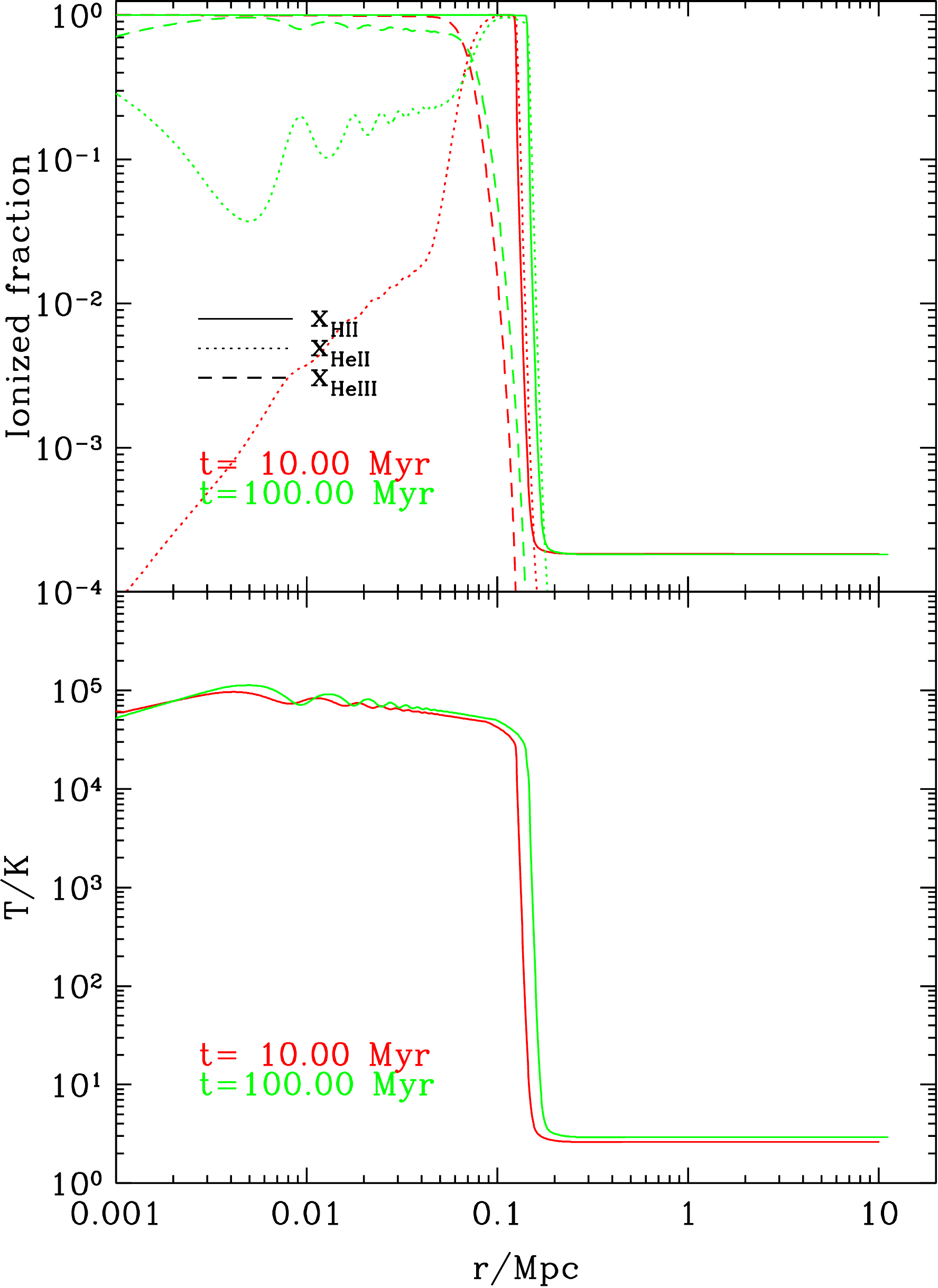} & 
\includegraphics[width=1.5in]{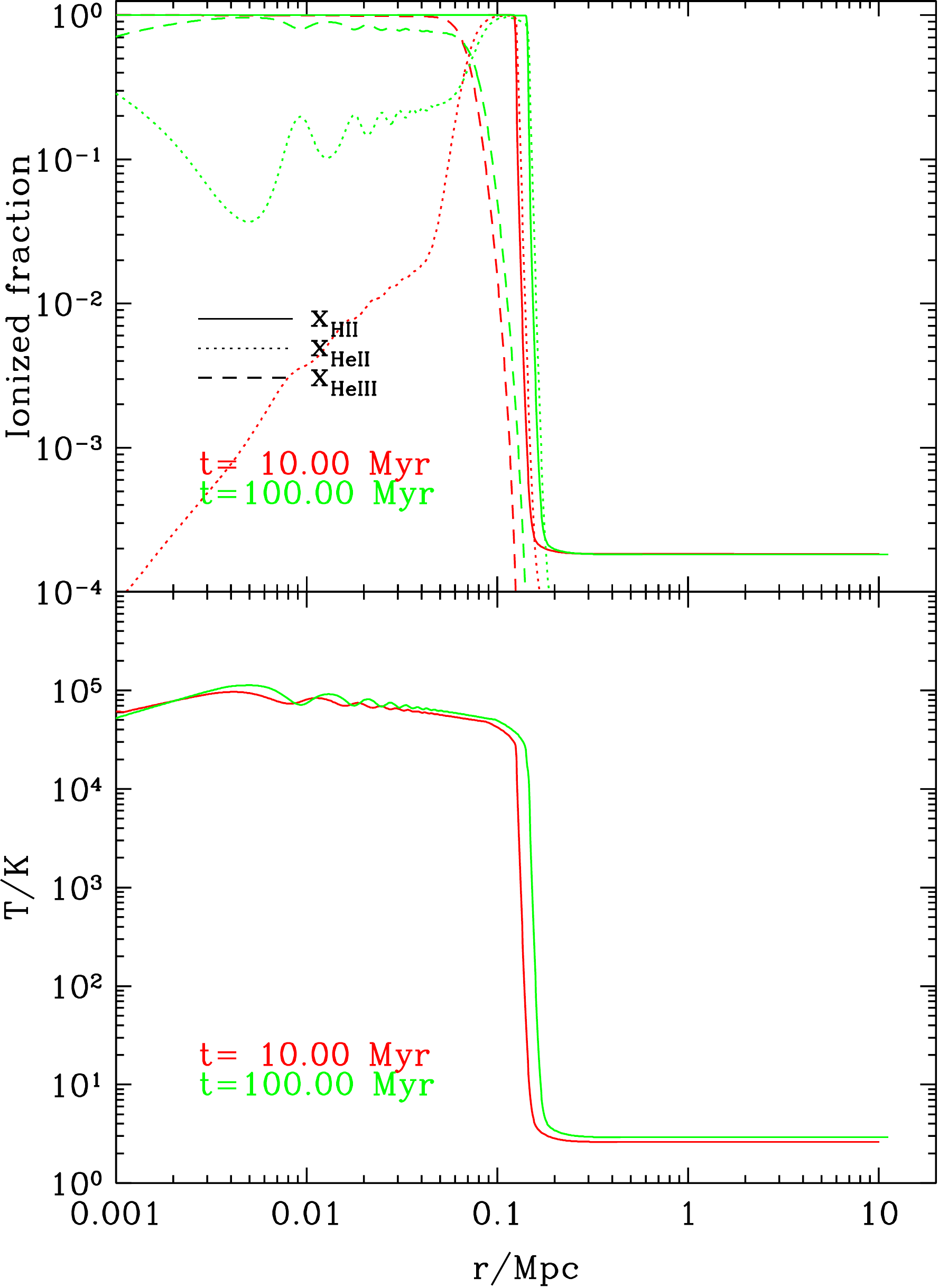} \\
\includegraphics[width=1.5in]{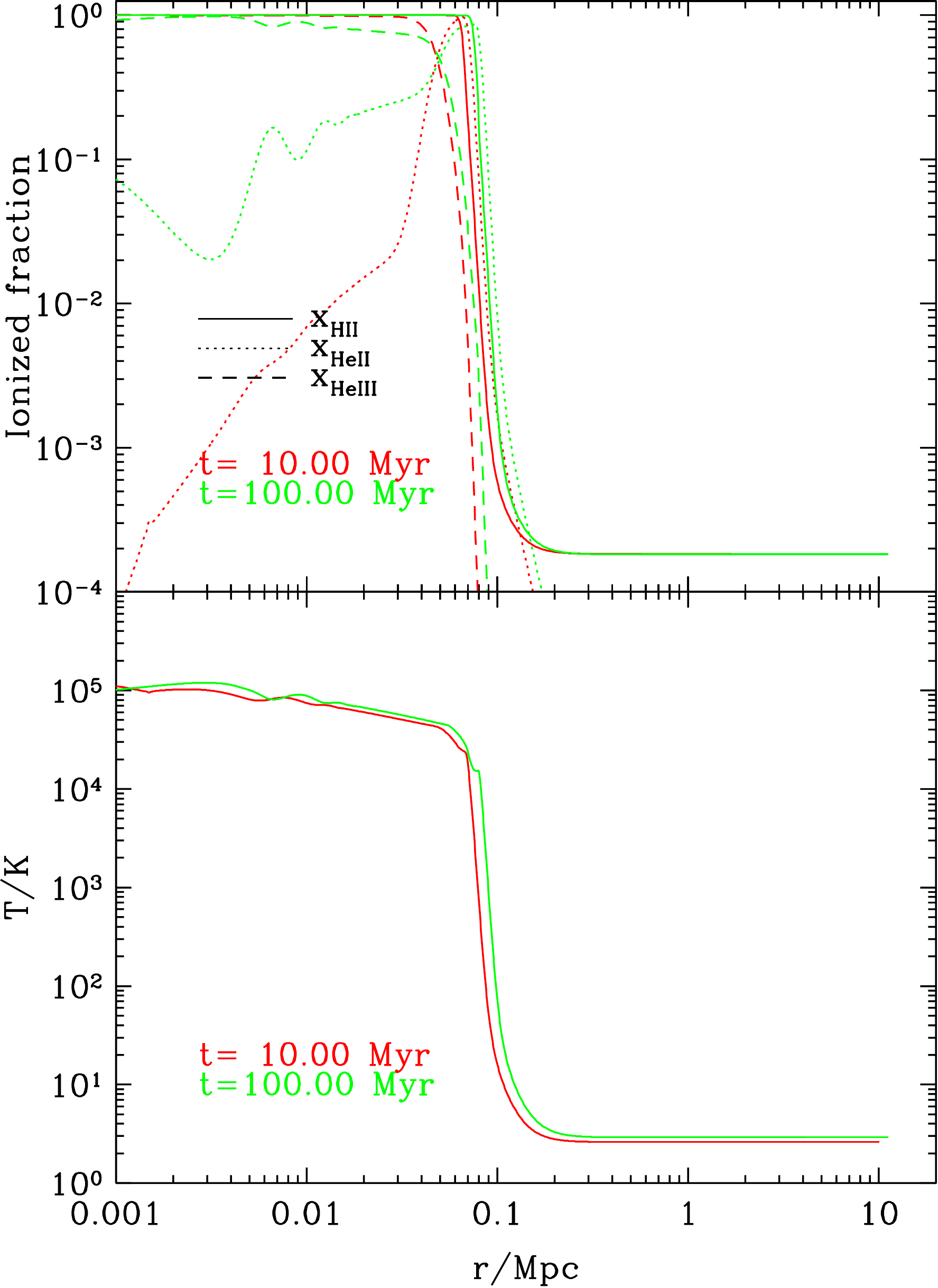} & \\
\end{array} $
\end{center}
\caption{The ionization and temperature profiles for a $10^6$ solar-mass starburst only (no QSO/BH) at  $z \sim 10$. }
\label{fig:stars}
\end{figure}

\begin{figure}
\begin{center} $
\begin{array}{cc} 
\includegraphics[width=1.5in]{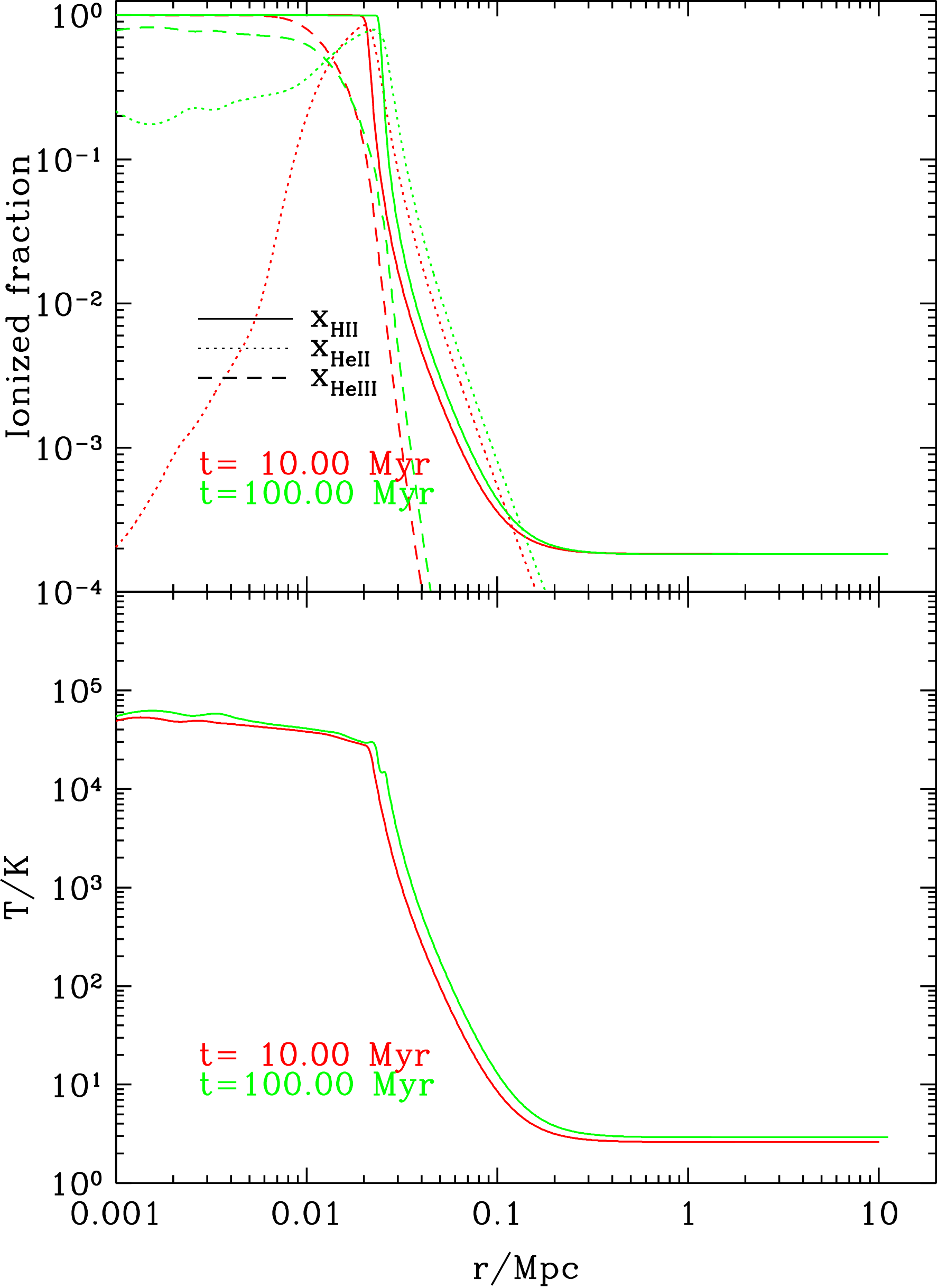} &
\includegraphics[width=1.5in]{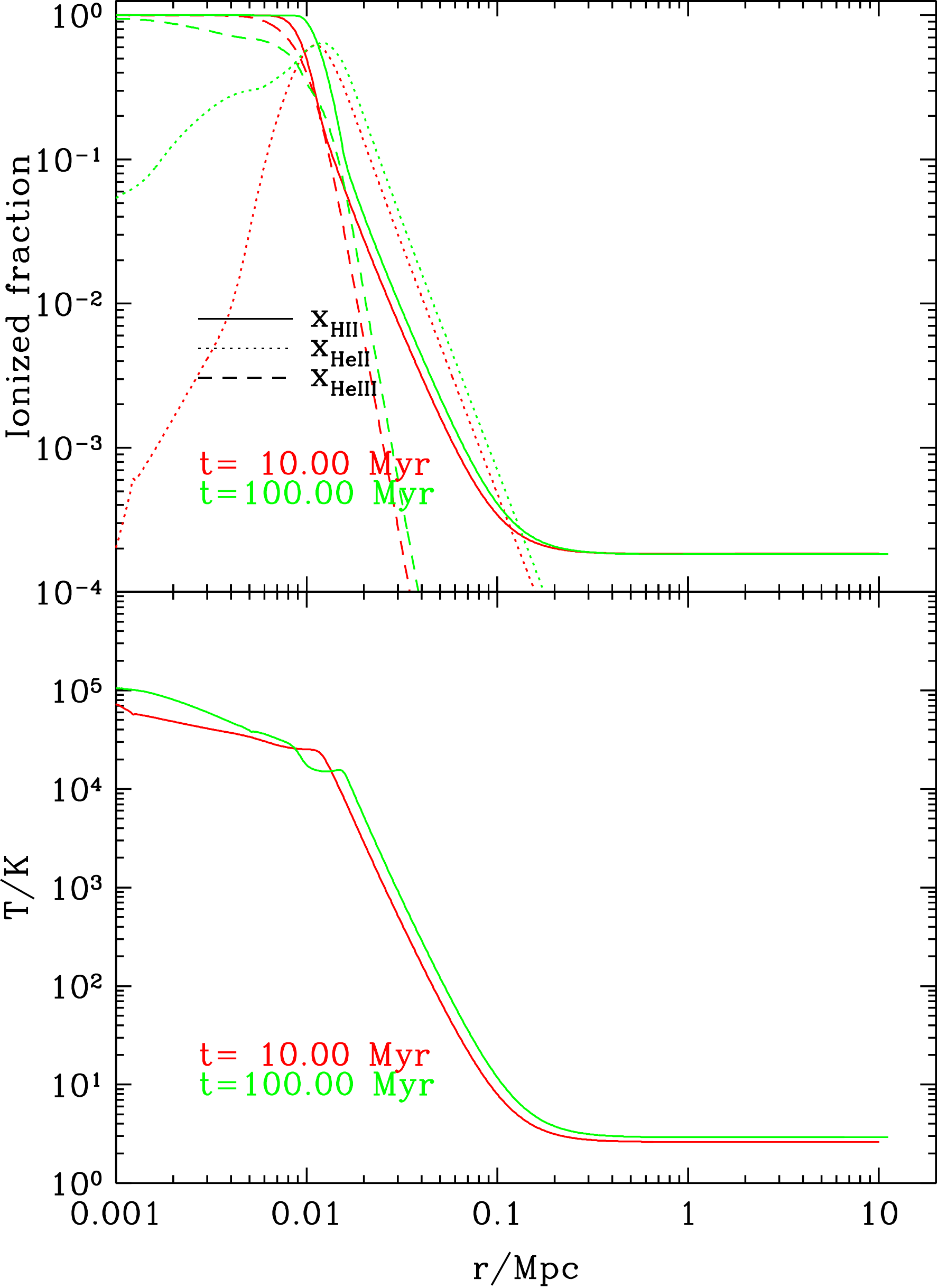}  \\
\includegraphics[width=1.5in]{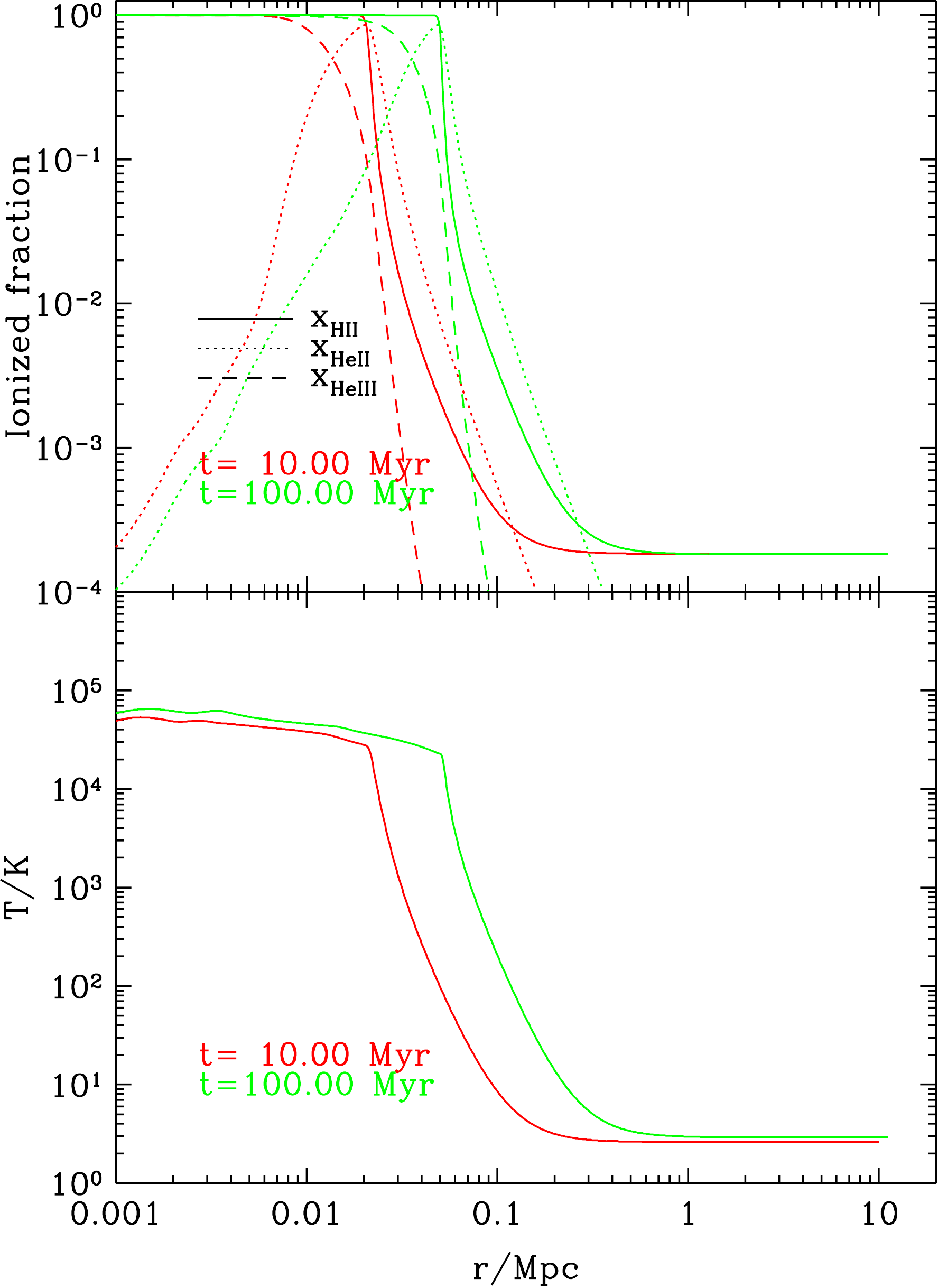} &
\includegraphics[width=1.5in]{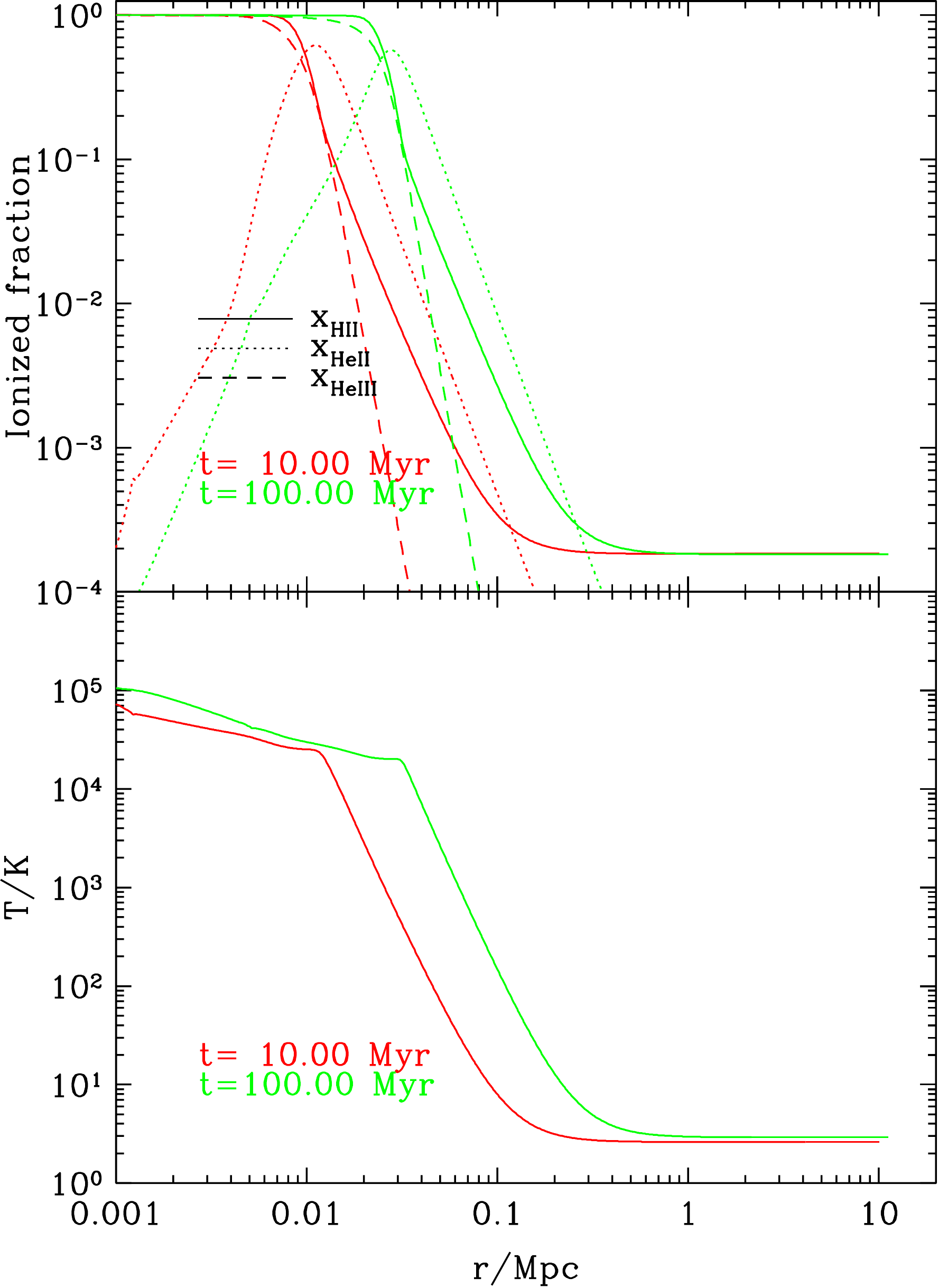}  \\
\end{array} $
\end{center}
\caption{The ionization and temperature profiles for a $10^3$ solar-mass starburst with $10^4$ solar-mass QSO/BH at  $z \sim 10$. The left and right panels in each row display the cases with the full QSO spectrum (including UV/X-ray photons) and with X-rays only. The upper row has a QSO duty cycle of 10 Myr, and the lower row has a  QSO duty cycle of 100 Myr. Red and green curves show the curves at times 10 Myr and 100 Myr after the quasar turns on.  Unlike previous figures in the paper, the no-Xrays case is not shown here, as it is very similar to the full spectrum case.}
\label{fig:smallbh}
\end{figure}

We also consider cases where the BH mass and QSO duty cycle are varied. This reveals the various contributions more clearly, particularly that from X-rays. The results are shown in Figures ~\ref{fig:bigbh}, ~\ref{fig:stars} and ~\ref{fig:smallbh}, where we can see that increasing (or decreasing) the BH mass or the duty cycle simply ``dials up" (or ``dials down") the effects of ionization. For the higher BH mass, the X-ray I-fronts advance further and reach higher values of ionization. Nevertheless, the high temperatures of $10^6$ K and strong ionization effects from X-rays  at large scales found by some authors, e.g. \citet{Thomas:08}, are not reproduced here, possibly arising from differences in model assumptions and input spectra (discussed further in Section 3.3).

Comparing the curves for the X-rays-only case  for QSO BH masses of 0,  $10^3$ M$_\odot$ and $10^8$ M$_\odot$, we see that X-rays {\it can} make a difference. Perhaps X-rays can become competitive with UV ionization only when the BH masses approach $10^8$ $M_\odot$. Note that such high QSO BH masses are very rare at $z=10$ (Figure ~\ref{fig:smbh}), and likely nonexistent at $z=20$ when the universe is younger and there has been little time to gain mass for a seed BH accreting at rates close to the Eddington value. Such $10^8$ M$_\odot$ or higher-mass AGN therefore may not contribute significantly to a cosmic X-ray background at $z \ga$ 10.
Also, we point out that in all the figures the X-ray related features noted earlier (the tail of low-level ionization in H~II and He~II, but not He~III, at large radii) are evident in the upper two panels in each case. The exception is the case with only stars ($10^6$ M$_\odot$ starburst, BH mass of 0) where the figures with and without X-rays are (unsurprisingly) near-identical.

Additionally, we ran cases with smaller masses in stars and BHs. One such case is shown in Figure ~\ref{fig:smallbh}, where the ionization and temperature profiles are displayed for a $10^3$ M$_\odot$ starburst with $10^4$ M$_\odot$ BH at  $z \sim 10$, at times 10 Myr and 100 Myr after the quasar turns on. Unlike previous figures in the paper, the no-Xrays case is not shown here, as it is very similar to the full spectrum case. The various panels show  the curves for the full QSO spectrum (including UV/X-ray photons) and with X-rays only, with varying QSO duty cycles of 10 Myr, and 100 Myr.  The ionization and maximum temperatures are lower over 10--100 kpc compared with our standard case but the role of X-rays for He~I ionization is more clearly seen here than in most our cases, particularly in the X-rays only panel for a QSO duty cycle of 100 Myr.

Other trends include variations with time or between species. Allowing the QSO/starburst source to be ``on" for 100 Myr advances the I-fronts for all cases and species relative to the curves for 10 Myr, as expected. The temperatures, however, increase noticeably at 100 Myr only for the pure X-rays case; for the cases involving the full spectrum or without X-rays, the temperatures appear to saturate at a few tens of thousands of degrees Kelvin, and having the source on for longer timescales makes little difference. In addition, 
the He~III I-front mostly lags the H I-front but in some cases the He~III front almost catches up to the H I-front. Thus, it appears that these species' I-fronts can be coincident for sufficiently hard radiation. 

The He~II ionization fraction exceeds that of H~I by a small margin, particularly beyond the edge of the UV I-front. We recognize this as the characteristic tail of added  secondary ionizations from X-rays, which manifest more strongly at larger physical scales where the UV photons do not penetrate as far. This can be seen best by comparing the no-X-rays and  X-rays-only panels of all the figures in this section, where the He~II front lags or is similar to the H~I front when X-rays are absent but leads the H~I front when only X-rays are present. This interplay between X-ray secondary ionization and the ionization balance of H and He in the presence of hard radiation leads to ionization boundaries  that are less sharp than in the UV-ionization case alone (see also \citealt{furlanetto08a} on this point in relation to the morphology of helium reionization at lower redshifts,  $z \sim$ 3).  Last, in the case with only a $10^6$ M$_\odot$ starburst (Figure ~\ref{fig:stars}), we see that there is little difference between these two panels, as this case has low X-ray production.

\begin{figure}
\begin{center} $
\begin{array}{cc} 
\includegraphics[width=1.5in]{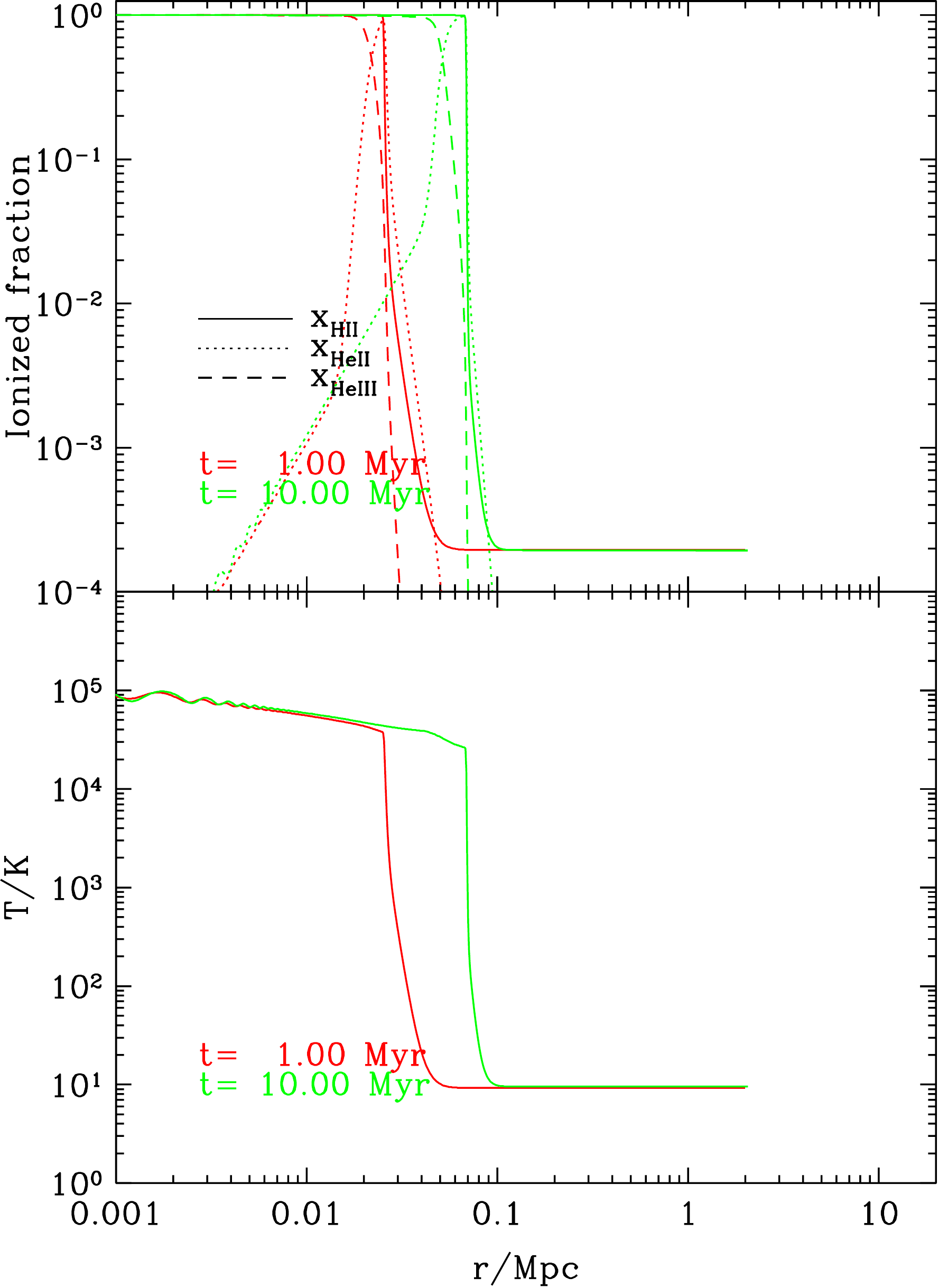} & 
\includegraphics[width=1.5in]{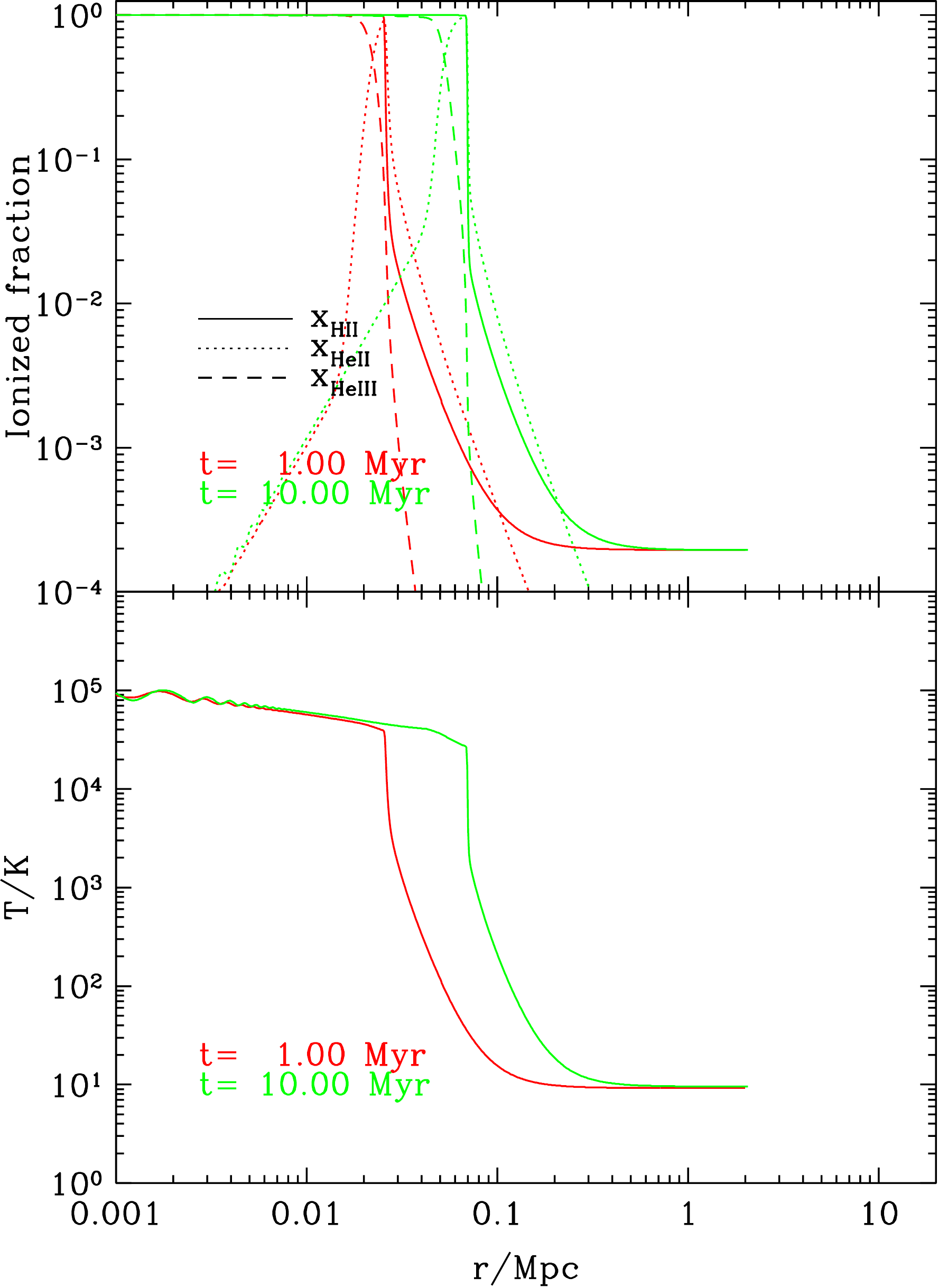} \\
\includegraphics[width=1.5in]{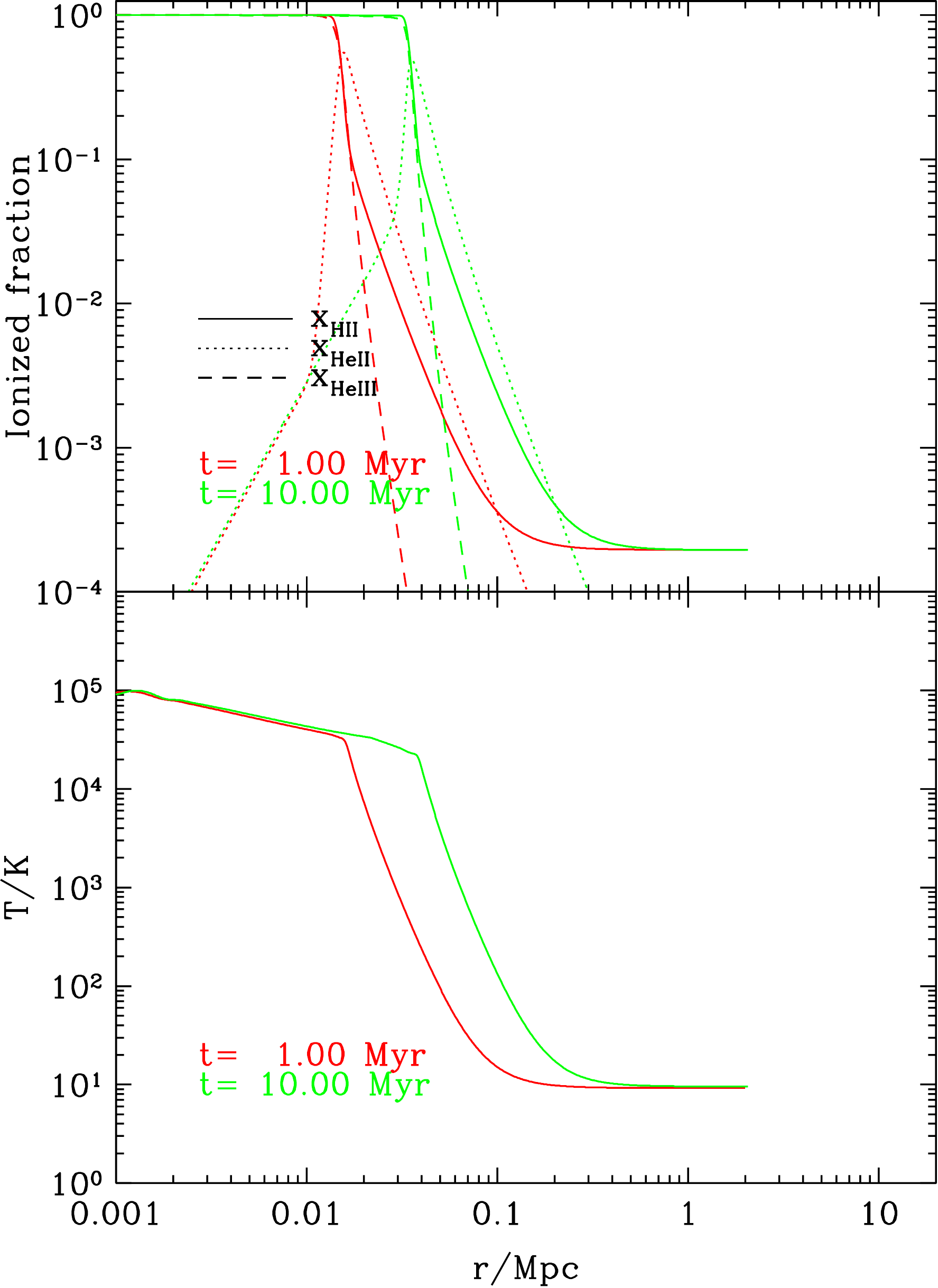} & \\
\end{array} $
\end{center}
\caption{The ionization and temperature profiles for a $10^5$ solar-mass starburst with $10^6$ solar-mass QSO/BH at  $z = 20$. The legend is the same as in earlier figures. See text for explanation. Curves are shown for timescales of 1 and 10 Myr (rather than 10 and 100 Myr as in all preceding figures), owing to the shorter IGM recombination timescales at $z=20$ relative to $z=10$.}
\label{fig:ionz20}
\end{figure}

To test the variation with redshift, we perform the same calculations for our standard case assumptions at $z=20$. Exploring redshifts lower than $z \sim$ 10 marks the era of  overlapping I-fronts as reionization draws to an end, which our current treatment cannot model well. Additionally, there is not much H~I remaining outside of galaxy halos to generate an interesting 21 cm signal at the end of reionization, whereas the 21 cm signal is expected to be significant at $z = 10$--20. The calculations at $z=20$ for our standard case are displayed in Figure ~\ref{fig:ionz20} with the same three panels as in the ionization and temperature figures. As the IGM is denser and the recombination timescales are shorter, we show curves for times at 1 Myr and 10 Myr (rather than 10 Myr and 100 Myr) after the source turns on. We see that the ionization curves at 10 Myr between the $z=20$ case and our standard case at $z=10$ have very similar shapes, with the $z=20$ curves lagging the $z=10$ curves, expected from the higher IGM densities at earlier times. Note however that the peak temperatures achieved in all of these cases remain similar, around $10^5$ K.

We perform a simple estimate of the tradeoff between the local X-ray flux from a single galaxy versus the X-rays from a number of distant sources. The comoving number density of halos in our work with masses $\ga 10^8$ $M_\odot$ is, $n =$ 1.147 (6.443 $\times 10^{-4})$ Mpc$^{-3}$ at $z=$10 (20). This translates to  an average spacing between such halos of $\sim$ 0.95 (11.5) Mpc at $z=$ 10 (20). The emission rate of H-ionizing photons for a $10^5$ M$_\odot$ starburst with a $10^6$ M$_\odot$ BH (our typical case)\footnote{For comparison, $S \sim 0.43 \times 10^{51}$ photons s$^{-1}$ for a single 200 M$_\odot$ star in \citet{chen08}, $S \sim 10^{52}$ photons s$^{-1}$ in \citet{ricotti05} (from the discussion related to their equation 4), and $S \sim 10^{50}$--$10^{54}$ photons s$^{-1}$ for the BH mass range of $10^3$--$10^6$ M$_\odot$ considered in \citet{Thomas:08}. } is $S \sim 1.3 \times 10^{51}$ photons s$^{-1}$. The associated X-ray photon production rate is $\sim 1.3 \times 10^{49}$ ($2.1 \times 10^{48}$) photons s$^{-1}$ at 300 eV and 1 keV respectively. If we assume a uniform IGM with no attenuation and that the visibility sphere for sources can go out to a maximum radius given by the MFP derived for X-rays as a function of energy in Fig. ~\ref{fig:mfp}, then the critical distance from an individual galaxy source at which the flux of the source become equal to the background flux from sources of similar individual fluxes is 0.1--0.5 Mpc at $z=10$ for 300 eV to 1 keV X-rays. Thus, our results at $z=10$, e.g. in Figure ~\ref{fig:ionstd}, could have additional contributions to X-ray ionization from neighboring galaxy halos at radii 0.1--1 Mpc, although this will be less of an issue at $z=20$.   In reality, we need to factor in realistic density profiles for the galaxies and the IGM, as well as the time variability of individual sources. We will pursue this in future work involving a full cosmological calculation through extensions to the current Galacticus code (see \S 4).

Last, we note the  oscillations in the He~II fraction and temperature profiles in some of our models. We performed a number of checks to make sure these were not 
mere numerical effects. We found that these oscillations are robust to increases in the time resolution, ODE solver
accuracy and number of radial shells used in our code. These oscillations are also well-resolved radially, 
and have a near-constant wavelength, despite the logarithmically-spaced grid spacing in radius.
What may be occurring is similar to the physics of the instability
strip in stellar atmospheres. Inside the ionized region, the optical depth is
very small, so the incident flux drops as $1/r^2$.  The small H~I, He~I and He~II
fractions are determined by the balance between photoionization, collisional
ionization and recombination rates, while the temperature is controlled by the
balance of photoheating and cooling rates. 
As we move outward in radius, this leads to a complex interplay between the photoheating rate,
temperature and the He~II fraction in the region of the He~III to He~II transition, leading  to the temperature and He~II fraction
oscillating with radius. This arises from our solving the time-dependent ionization and heating equations rather than adopting the
equilibrium solution. Given several of our idealized approximations here such as spherical symmetry, we do not expect this effect to have a significant impact, particularly  on the 21 cm signal which we discuss next.

\subsection{Radio Signatures}

Over the last decade, there has been a growing literature on the 21 cm radio signals arising from the percolation of reionization, i.e., the growth of ionized bubbles around the first luminous sources and the associated heating \citep{zaldarriaga04,chen04,chen08,kuhlen06,McQuinn:06,furlanetto04,furlanetto06,furlanetto06a,pritchard07,Thomas:08,Ripamonti:08,santos08,morales10}. The signature is expected to be absorption (emission) against the CMB if the ionized region is colder (warmer) than the CMB at those epochs. Forthcoming interferometric experiments at radio wavelengths, such as LOFAR and SKA, are predicted to be able to resolve ionized bubbles of size $\sim$ 100 kpc up to a few Mpc. The dominant signal arises from the coupling of the spin temperature of neutral hydrogen with the kinetic temperature of the background IGM gas. After recombination, the IGM cools as (1 + $z$)$^2$ whereas the CMB cools as (1 + $z$), leading to a 21 cm absorption signal from the neutral IGM gas. At later epochs, the spin states of hydrogen come into equilibrium with the CMB, leading to a decreasing 21 cm signal. As the first stars and quasars turn on, a 21 cm emission signal is generated through coupling the spin states with the scattering of Ly$\alpha$ photons and other processes.

Here, we follow the formalism outlined in \citet{chen08}. As we do not follow the detailed cosmological evolution of a distribution of ionized bubbles, we model the spin temperature of H~I at a fixed redshift as: \\ 
\begin{equation}
T_s = \frac{T_{\rm CMB} + (y_\alpha + y_c) T_k}{1 + y_\alpha + y_c}
\end{equation}

\noindent where $T_{\rm CMB}$ is the CMB temperature at that redshift ($z=10$ in our cases unless otherwise specified) and $T_k$ is the gas kinetic temperature (which is a function of distance from the source). The $y$-coefficients are related to the coupling arising from Ly$\alpha$ photons ($y_\alpha$) and from collisions ($y_c$). The coefficient $y_c$ is taken from \citet{chen08} and \citet{kuhlen06}.
The coefficient $y_\alpha$ is the Ly$\alpha$ coupling term arising from the Wouthysen-Field effect. We use the expressions for $y_\alpha$ from \citet{chen08}, \citet{zaldarriaga04}, and \citet{pritchard07}, with additional parameters from \citet{hirata06}.  In the cases considered here, Ly$\alpha$ coupling dominates over other terms such as collisional coupling. We specifically include the Ly$\alpha$ photons from the stars and/or QSO emission in our models, as well as the auxiliary Ly$\alpha$ photons arising from X-ray ionization \citep{chen08, Venkatesan:01, shull85}.

This leads to a brightness temperature (measured as a differential from the background CMB temperature at that epoch) given by: \\ 
\begin{equation}
\delta T_b = 40 \; {\rm mK} \, \; \frac{\Omega_b h_0}{0.03} \; \sqrt{\frac{0.3}{\Omega_0}} \; \sqrt{\frac{1 + z}{25}} \; \frac{\rho_{\rm H I}}{\bar{\rho}_{\rm H}} \; \frac{T_s - T_{\rm CMB}}{T_s}
\end{equation}

\noindent When this calculated brightness temperature, $\delta T_b$, lies above the CMB temperature at that epoch, the ionized region will be seen in emission against the CMB. Conversely, regions beyond the I-front that lie below the CMB temperature will be seen in absorption against the CMB.

\begin{figure}
\begin{center} $
\begin{array}{cc} 
\includegraphics[width=1.5in]{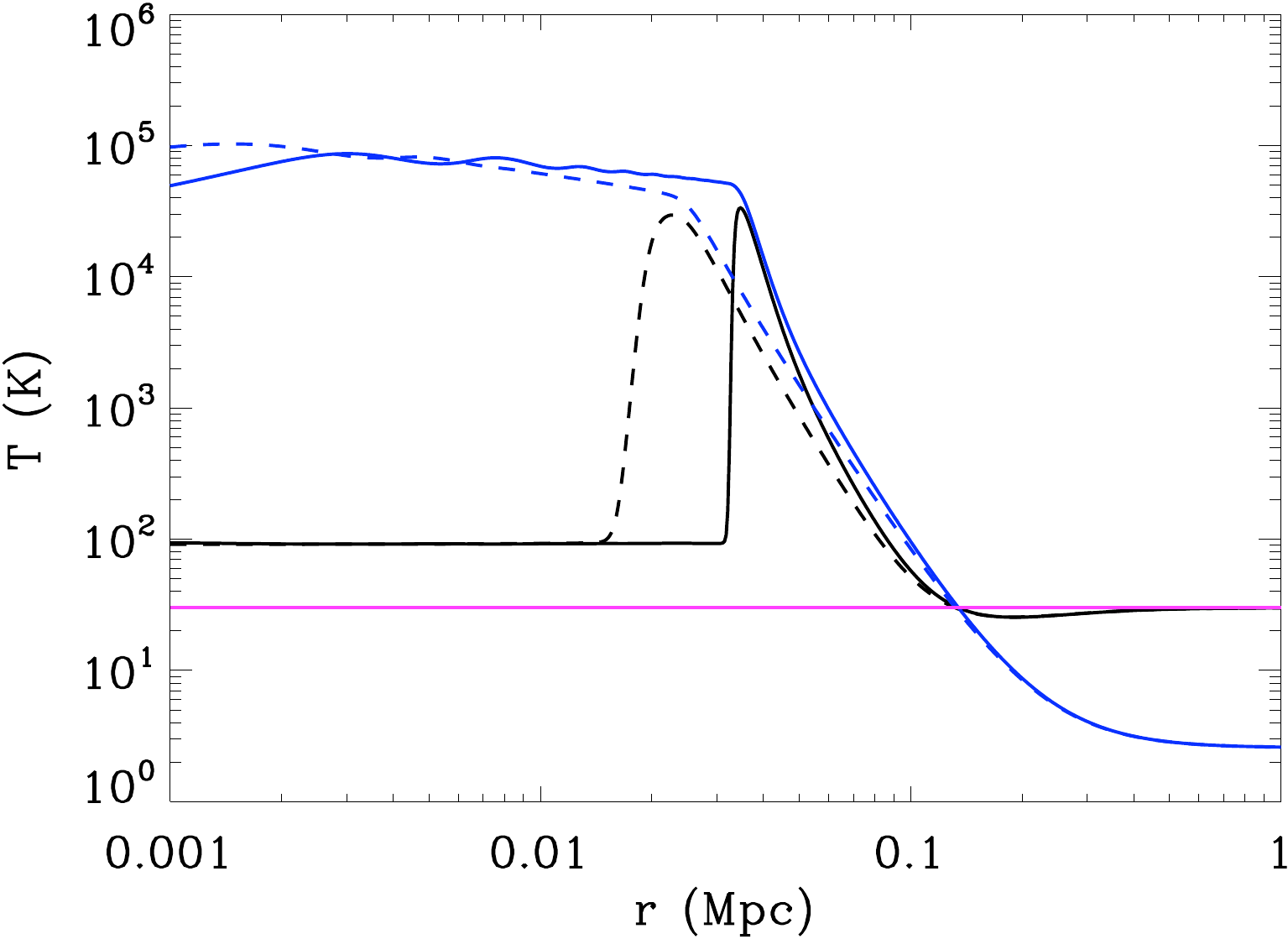} &\includegraphics[width=1.5in]{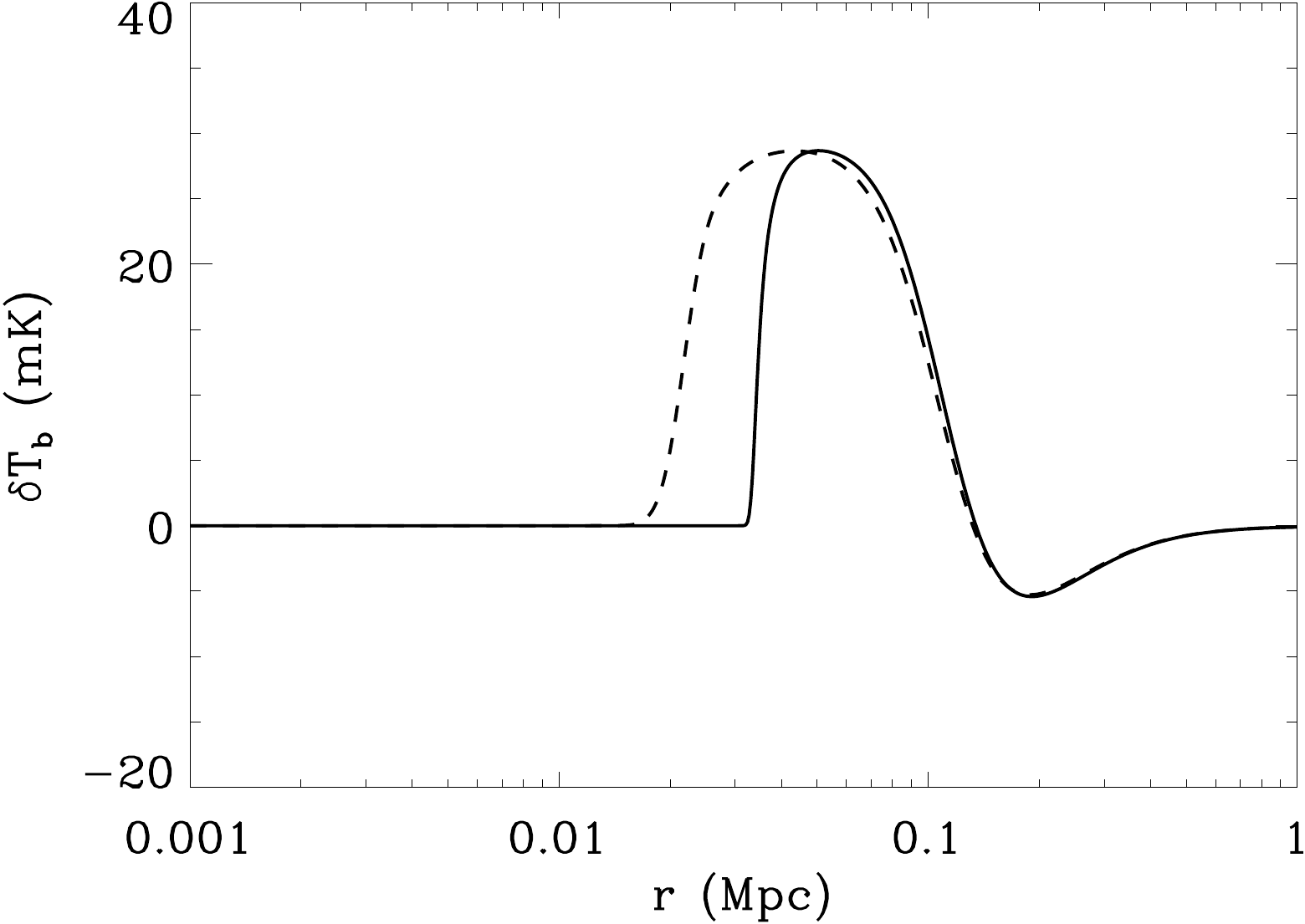}  \end{array} $\end{center}
\caption{Left panel shows the temperature profiles with radius for the spin temperature (black curves), kinetic temperature (blue curves) and the CMB temperature (purple line). Right panel displays the 21 cm brightness temperature profile. All cases are for a $10^5$ solar-mass starburst with $10^6$ solar-mass QSO/BH at  $z = 10$, at times of 1 Myr after the burst/QSO turn on. In each case, the solid lines are for the full spectrum case (X-rays and UV radiation) and the dashed lines are for X-rays only.  }
\label{fig:radiostd}\end{figure}

\begin{figure}
\begin{center} $\begin{array}{cc} 
\includegraphics[width=1.5in]{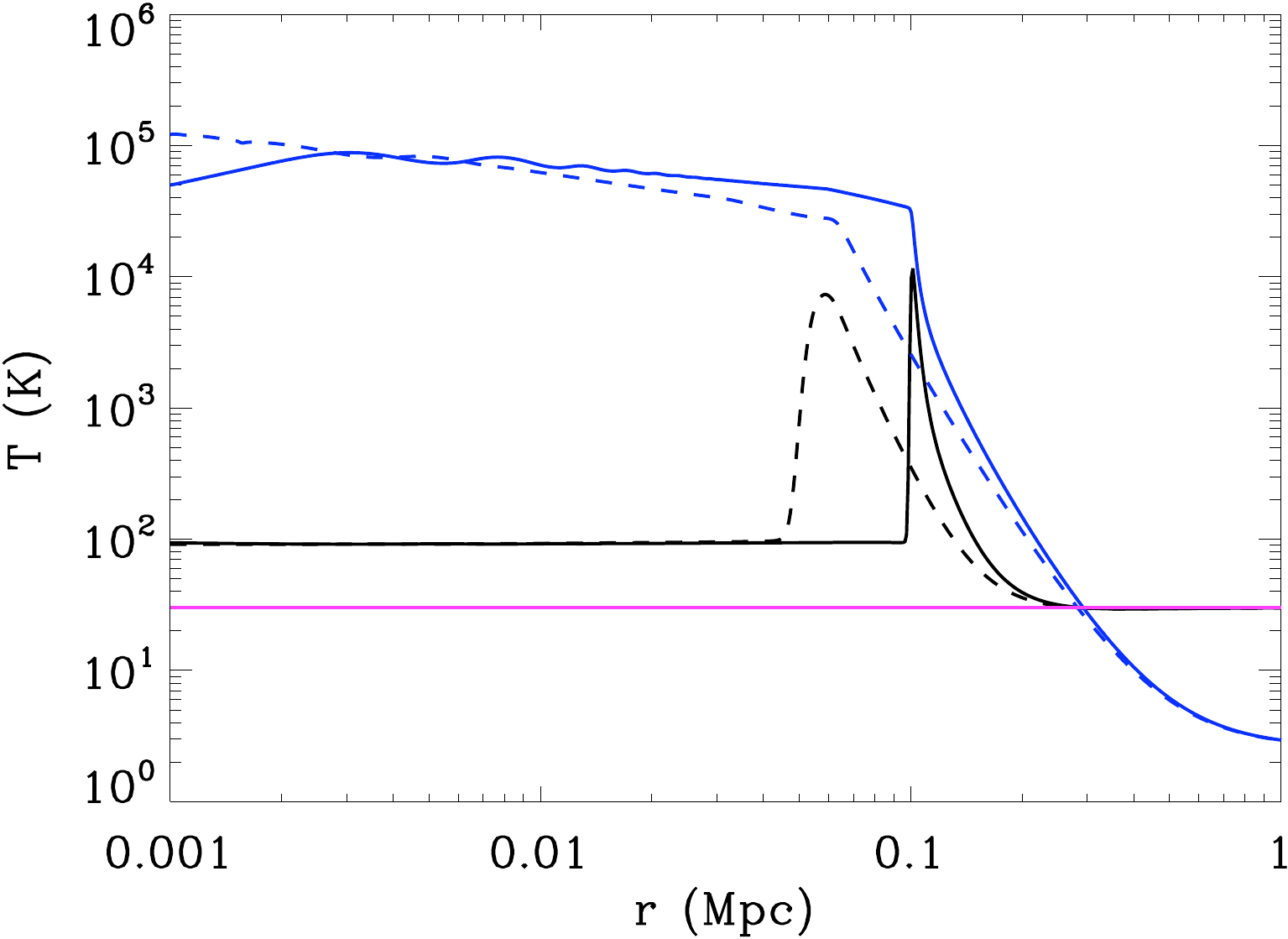} &\includegraphics[width=1.5in]{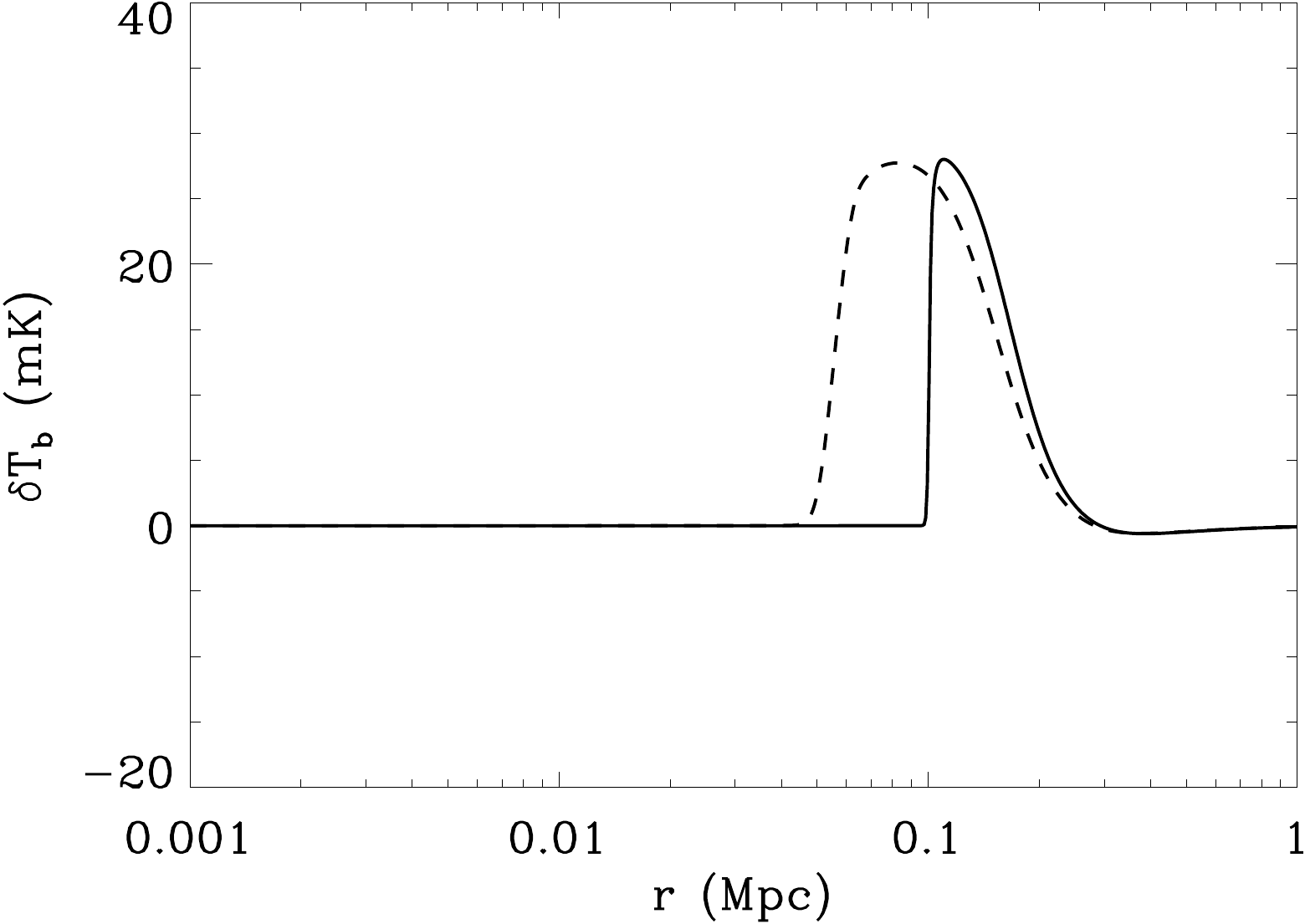}  \end{array} $
\end{center}
\caption{The temperature profiles (left panel) and 21 cm brightness temperature (right panel) are shown for a $10^5$ solar-mass starburst with $10^6$ solar-mass QSO/BH at  $z = 10$, at times of 10 Myr (rather than 1 Myr) after the burst/QSO turn on. The legend is the same as for Figure ~\ref{fig:radiostd}. }
\label{fig:radiostd10Myr}
\end{figure}

\begin{figure}
\begin{center} $\begin{array}{cc} \includegraphics[width=1.5in]{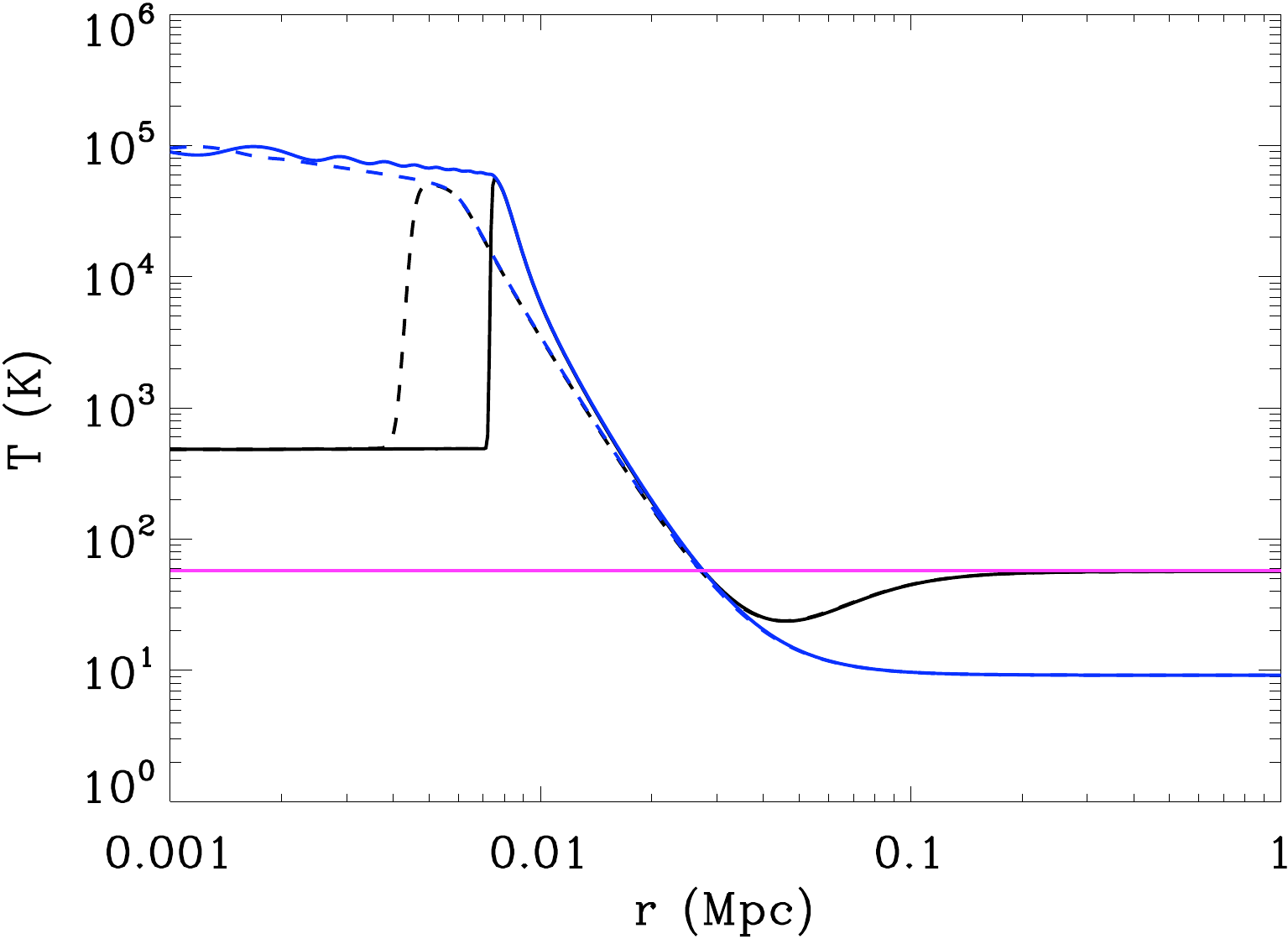} &
\includegraphics[width=1.5in]{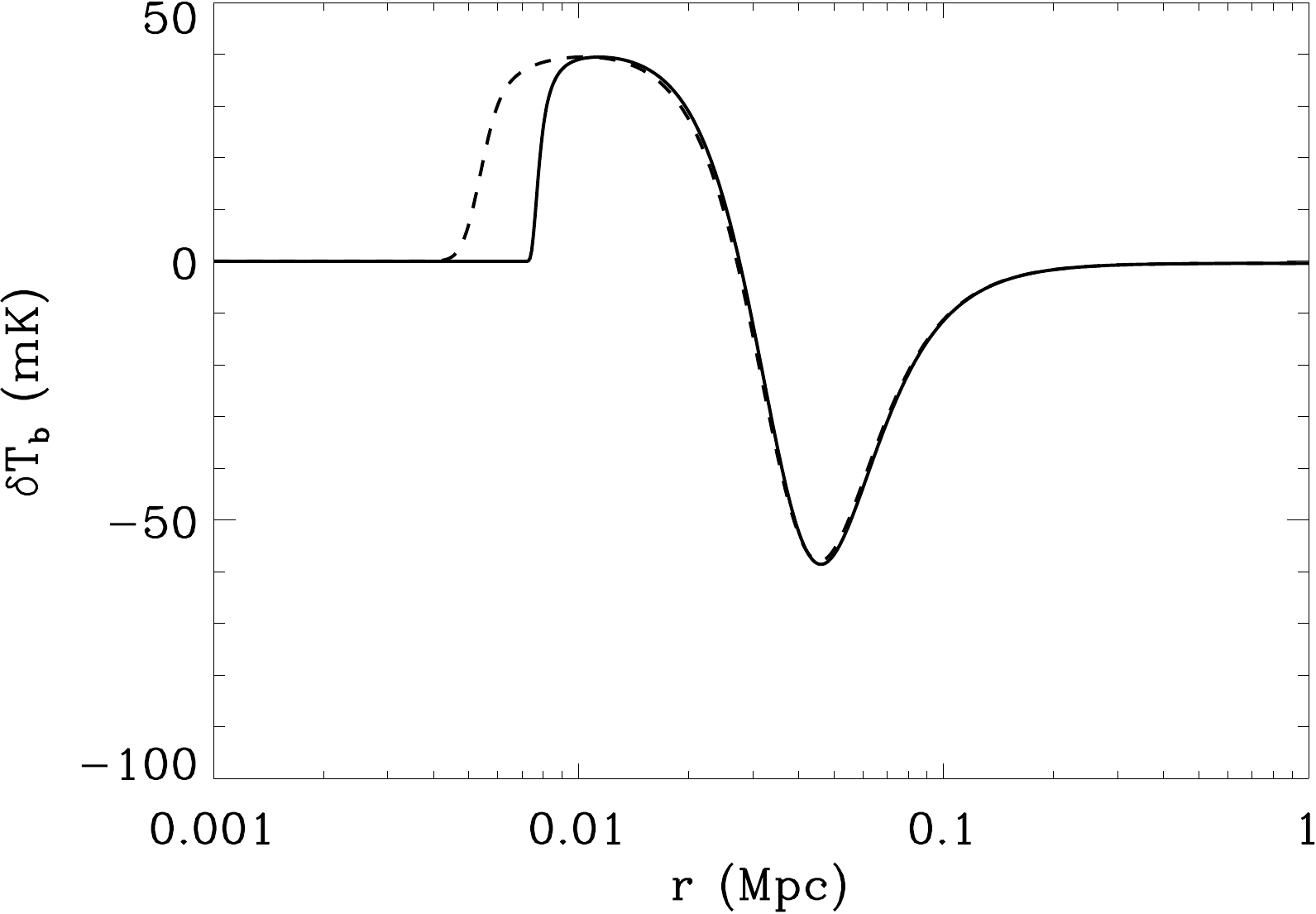}  
\end{array} $
\end{center}
\caption{The temperature profiles (left panel) and 21 cm brightness temperature (right panel) are shown for a $10^5$ solar-mass starburst with $10^6$ solar-mass QSO/BH at  $z = 20$, at times of 0.1 Myr after the burst/QSO turn on. The legend is the same as for Figure ~\ref{fig:radiostd}. Note the deeper trough in $\delta T_b$ (relative to the same case at $z = 10$)  in the absorption signal against the CMB at scales of $\sim$ a few tens of kpc. }
\label{fig:radiostdz20}\end{figure}

\begin{figure}
\begin{center} $
\begin{array}{cc} 
\includegraphics[width=1.5in]{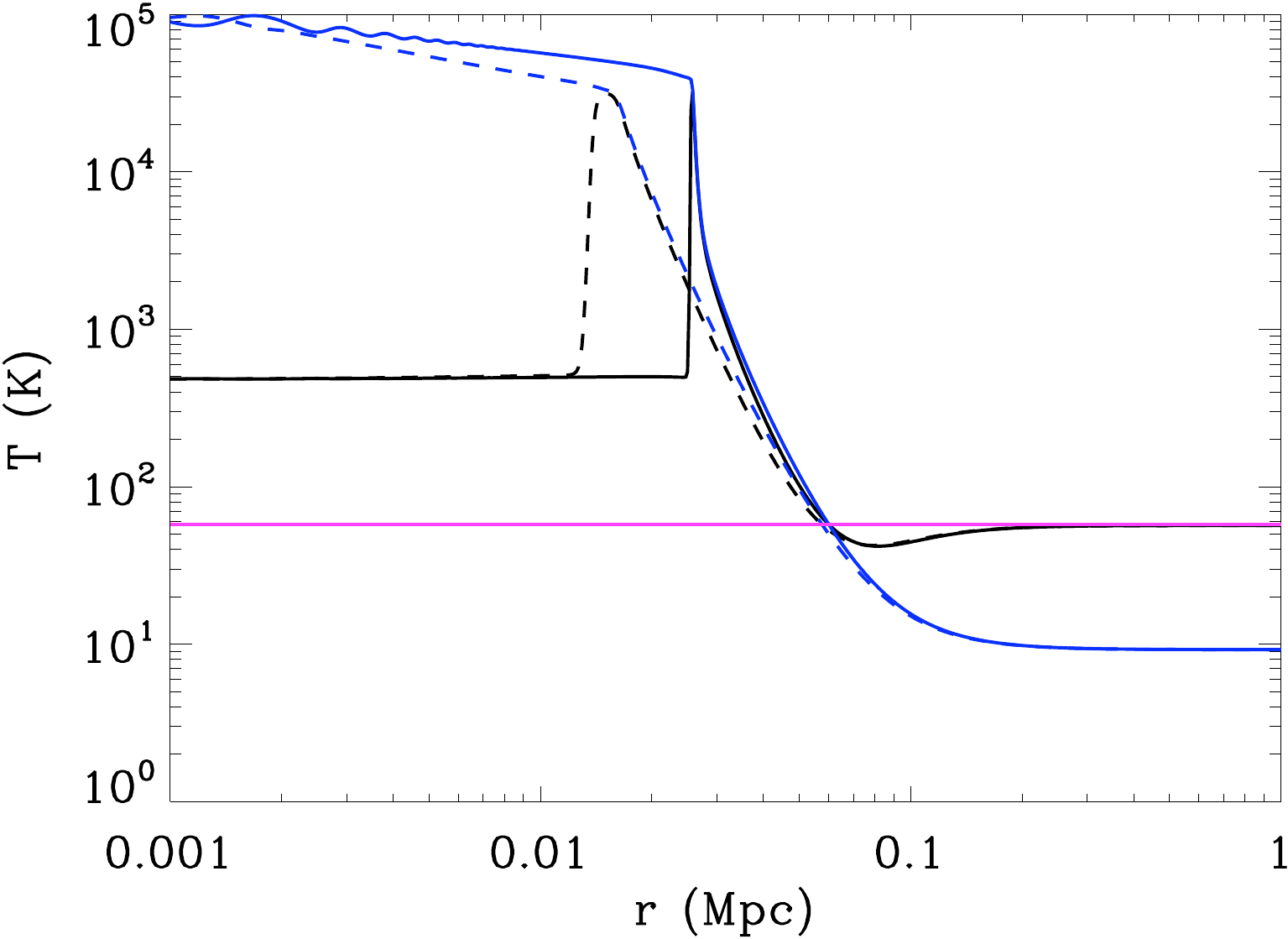} &
\includegraphics[width=1.5in]{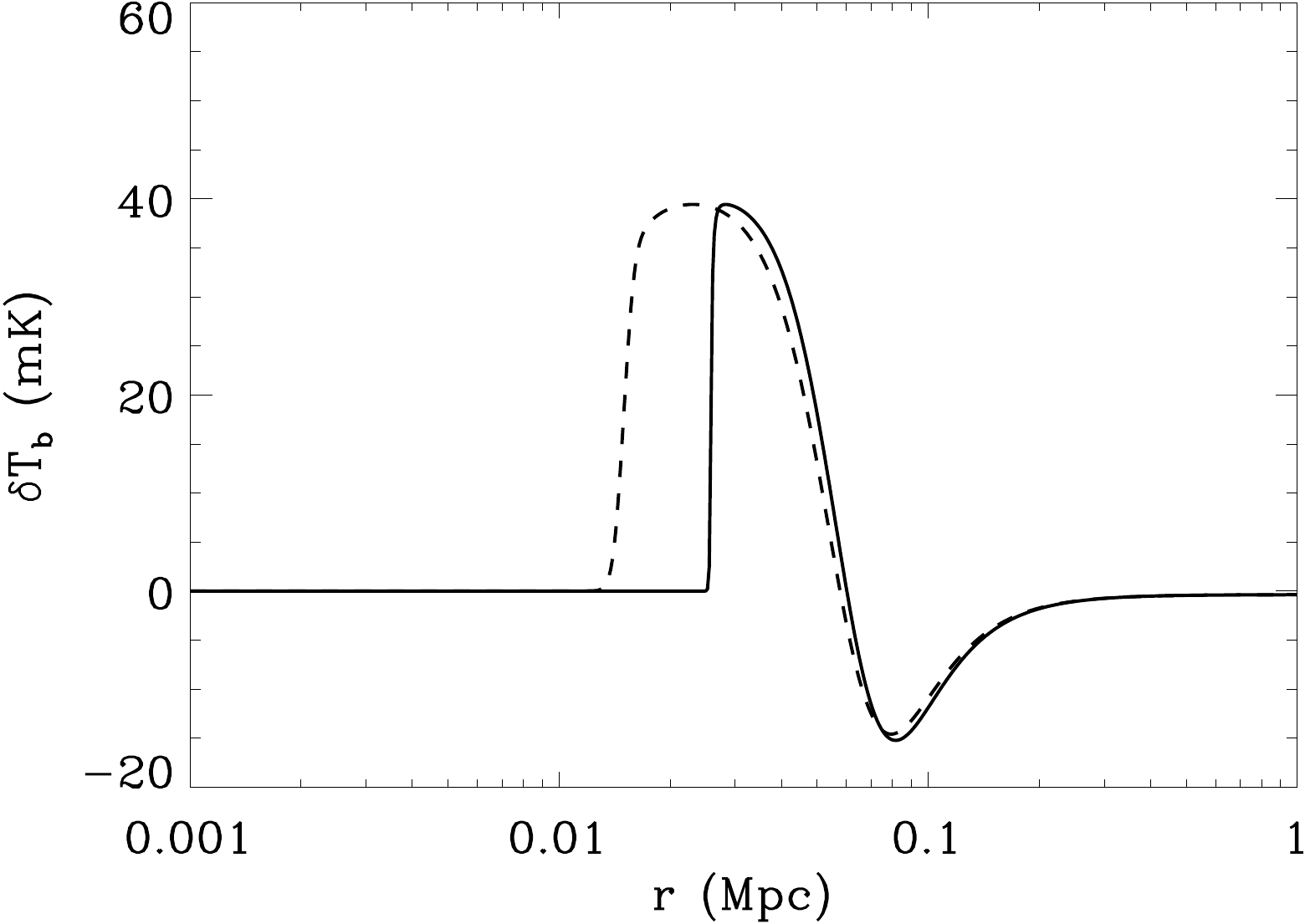}  
\end{array} $
\end{center}
\caption{The temperature profiles (left panel) and 21 cm brightness temperature (right panel) are shown for a $10^5$ solar-mass starburst with $10^6$ solar-mass QSO/BH at  $z = 20$, at times of 1 Myr (rather than 0.1 Myr) after the burst/QSO turn on. The legend is the same as for Figure ~\ref{fig:radiostd}. }
\label{fig:radiostdz20-1Myr}
\end{figure}

\begin{figure}
\begin{center} $
\begin{array}{cc}
\includegraphics[width=1.5in]{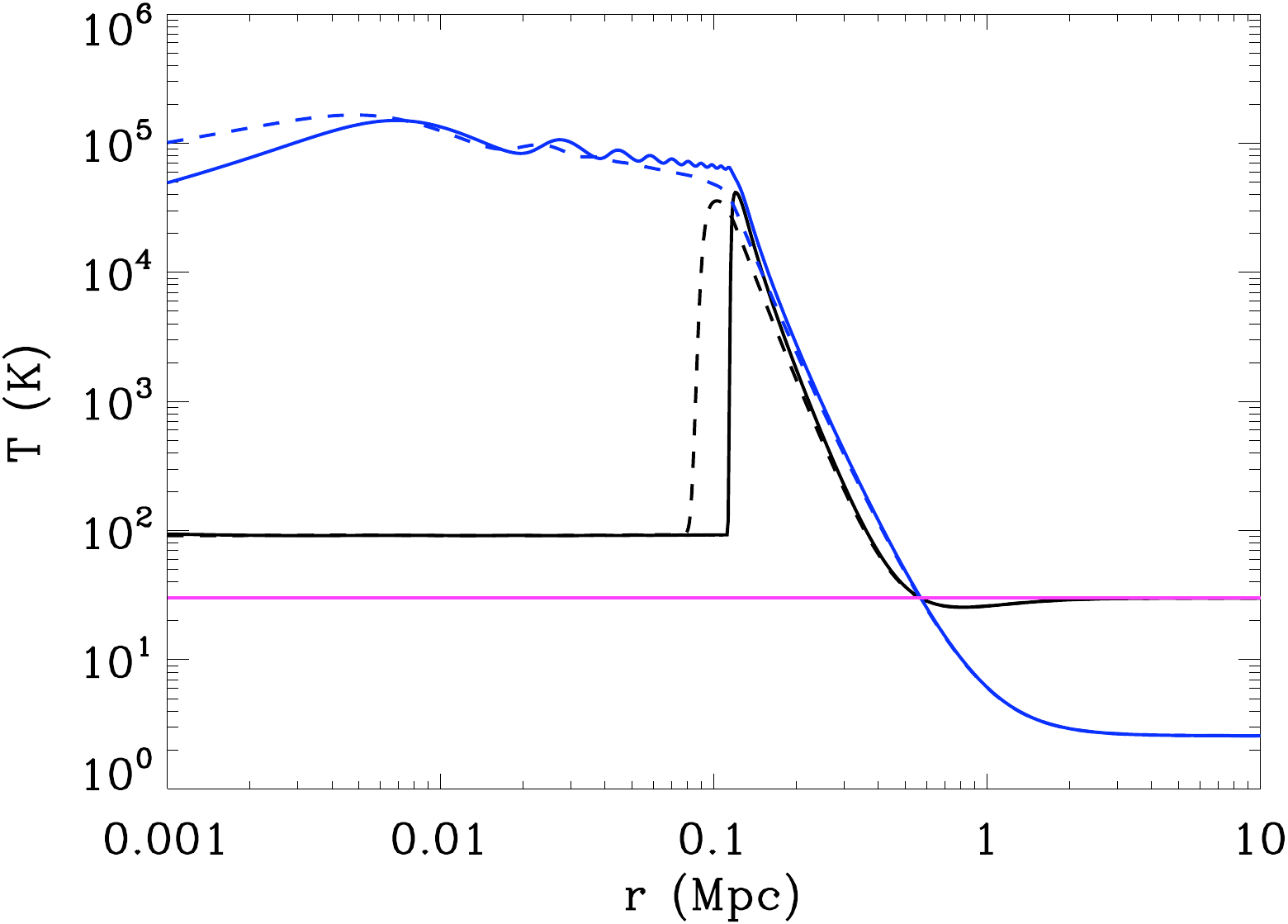} &
\includegraphics[width=1.5in]{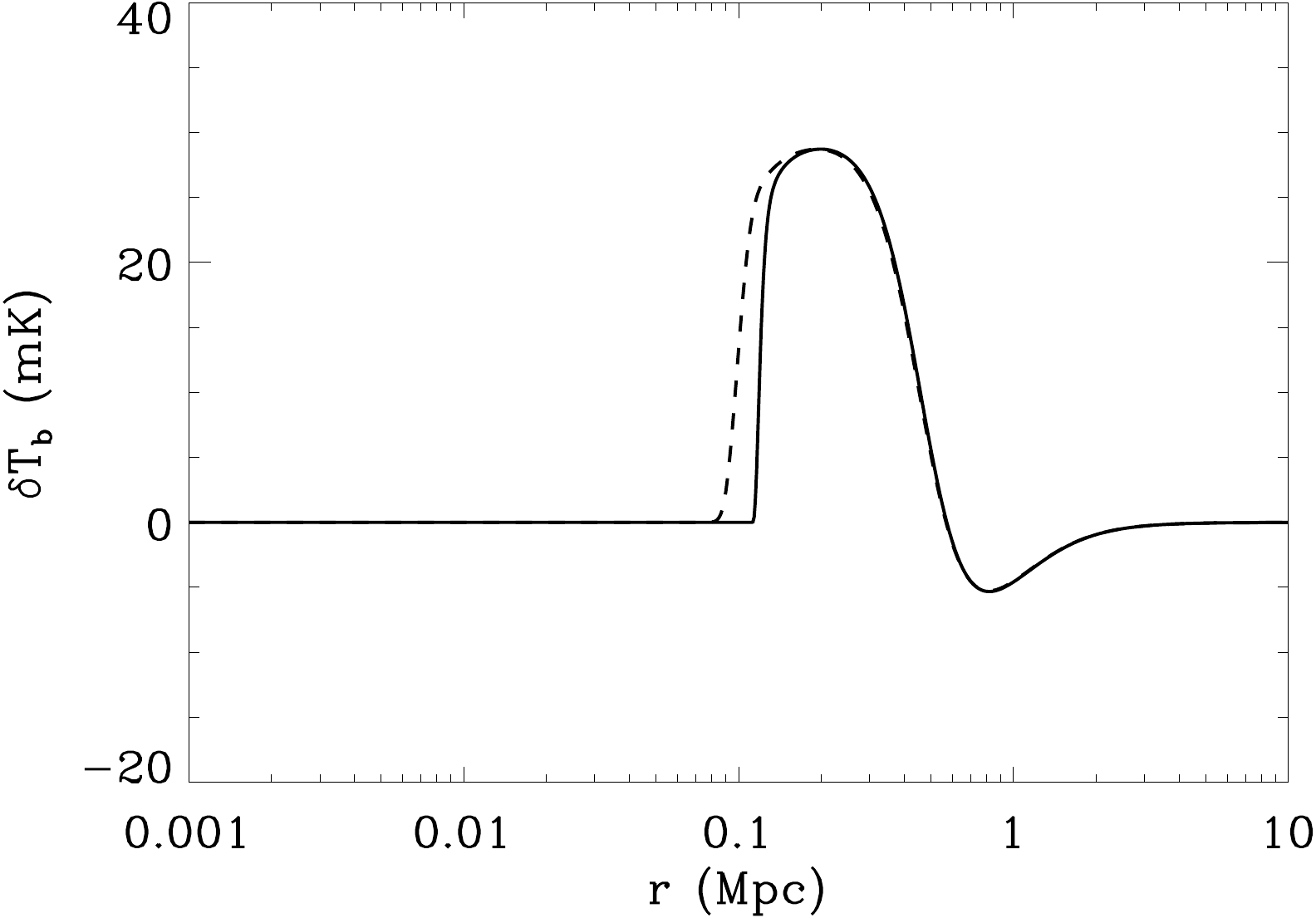}  
\end{array} $
\end{center}
\caption{The temperature profiles (left panel) and 21 cm brightness temperature (right panel) are shown for a $10^5$ solar-mass starburst with $10^8$ solar-mass QSO/BH at  $z = 10$, at times of 1 Myr after the burst/QSO turn on. The legend is the same as for Figure ~\ref{fig:radiostd}.}
\label{fig:radiobigBH}
\end{figure}

\begin{figure}
\begin{center} $
\begin{array}{cc}
\includegraphics[width=1.5in]{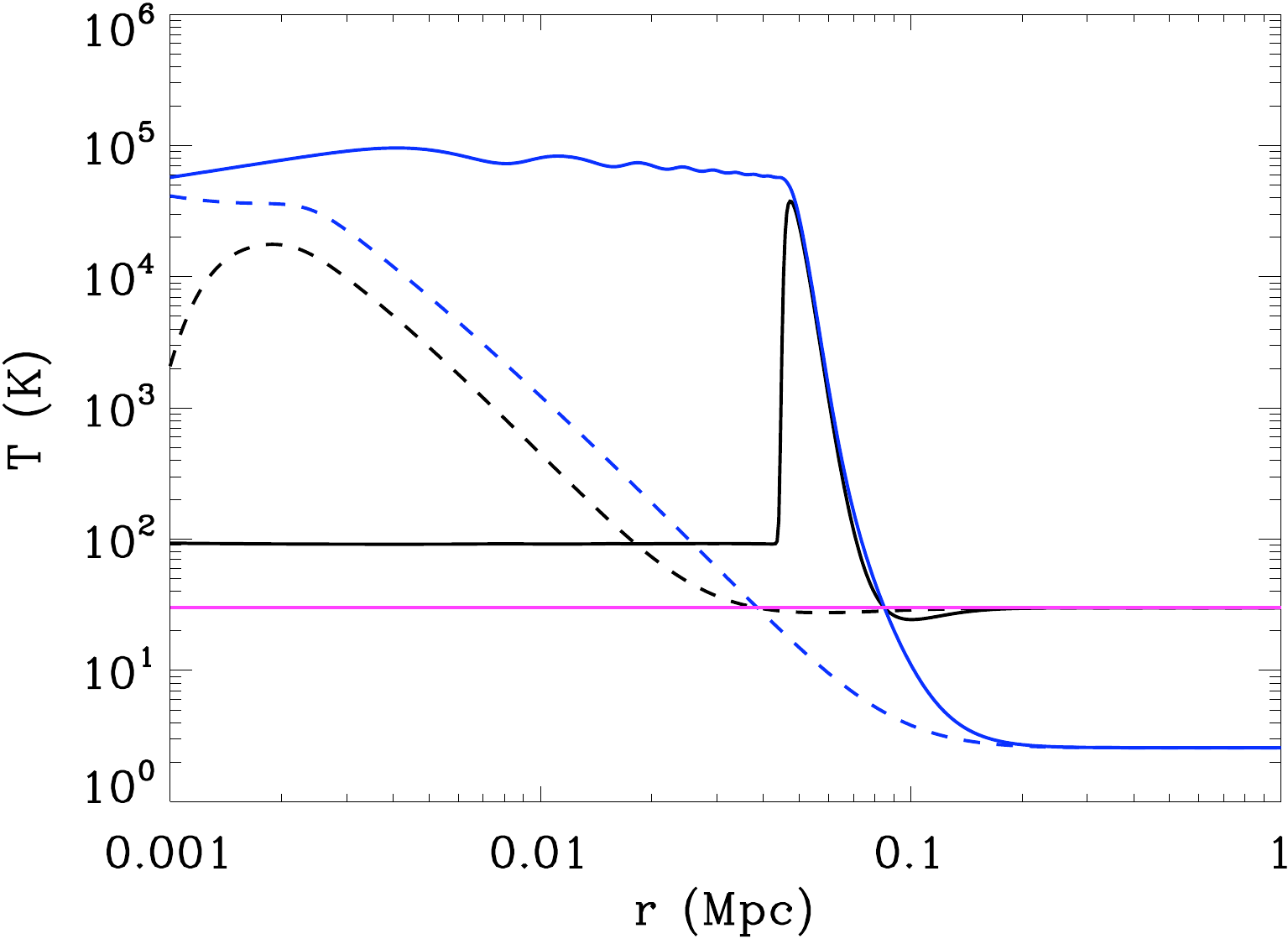} &
\includegraphics[width=1.5in]{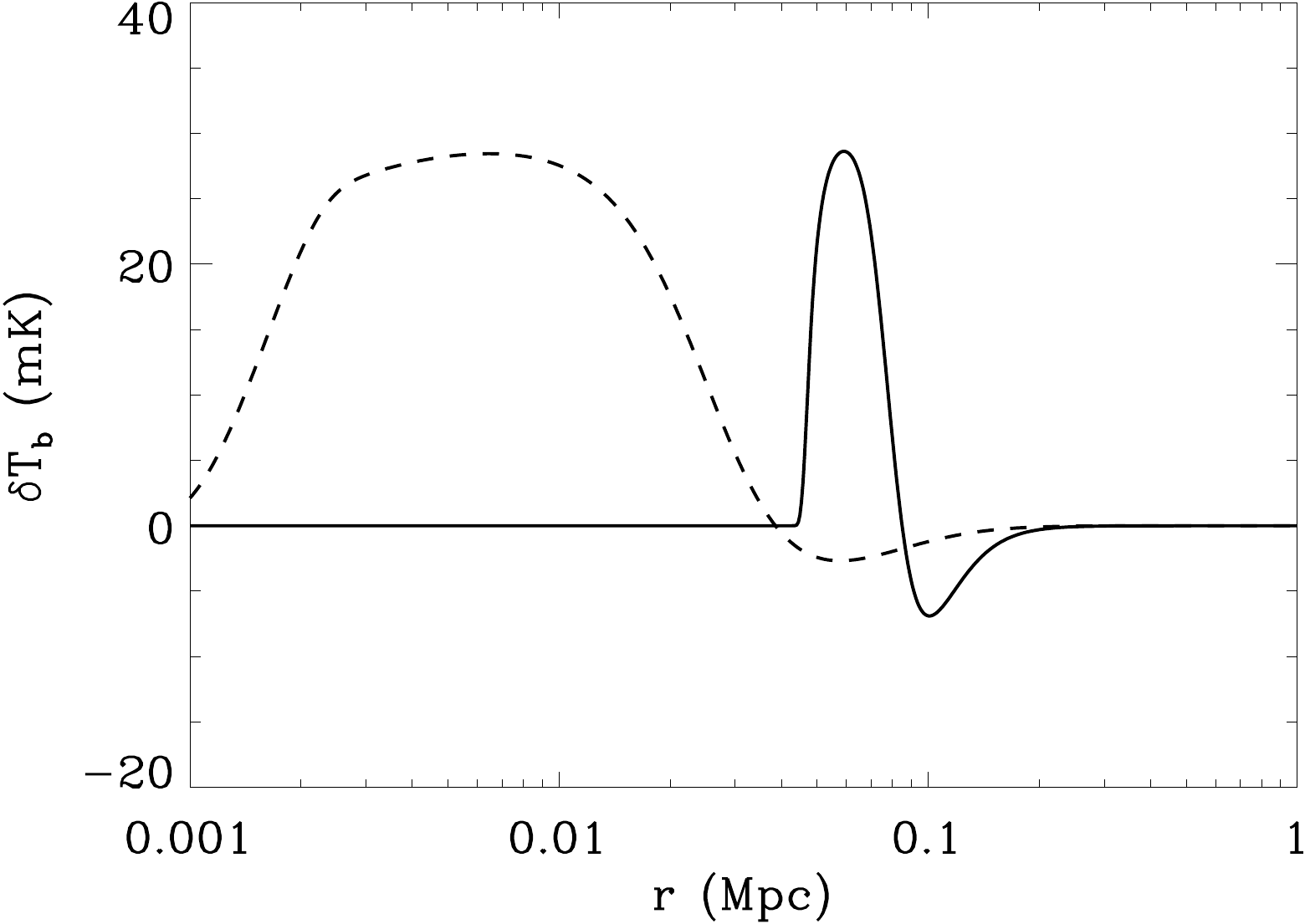}  
\end{array} $
\end{center}
\caption{The temperature profiles (left panel) and 21 cm brightness temperature (right panel) are shown for a $10^6$ solar-mass starburst with no QSO/BH at  $z = 10$, at times of 1 Myr  after the burst turns on. The legend is the same as for Figure ~\ref{fig:radiostd}. The difference is more pronounced between the cases with only X-rays versus X-rays and UV radiation, owing to the low X-ray production of stars in our models.}
\label{fig:radiostars}
\end{figure}

In Figures ~\ref{fig:radiostd}--~\ref{fig:radiostars}, we show the temperature profiles with radius for the spin temperature and gas kinetic temperature relative to the CMB temperature  which is constant at a fixed redshift. We also show the 21 cm brightness temperature profile and include the full spectrum case (X-rays and UV photons) and X-rays-only cases for each set of curves. These scenarios span most of the cases discussed in Section 3.1 involving a combination of starburst and QSO/BH masses (most of which are at $z=10$, with two cases at $z=20$).

Some broad conclusions that are common to all the cases whose 21 cm signatures are shown are as follows. First, the curves for the spin temperature are characteristically peaked around the location of the stalled I-front. The  transition from fully ionized within (with zero $\delta T_b$) to the neutral IGM gas occurs beyond the I-front in each case,  
with peak values for $T_s$ reaching $\sim$ 10$^4$--10$^5$ K in our cases, and peak values for the $\delta T_b$ emission signal around 30--40 mK. Negative $\delta T_b$ values, corresponding to an absorption signal relative to the CMB, occur on scales between 0.1 and 1 Mpc at $z=10$ in our models and have low net values of $\sim$ 0 to a few mK,  and larger values of $\sim$ 20--60 mK on scales of 0.01--0.1 Mpc at $z=20$.  We discuss this further below.

Second, the curves in each case corresponding to the X-rays only case for each starburst/BH scenario consistently {\it lag} the curves for the corresponding full spectrum case. This is most dramatically seen in the stars-only case (Figure \ref{fig:radiostars}), a $10^6$ solar-mass starburst with no QSO/BH), where the X-ray production is low. Here, the maximum values of $\delta T_b$ occur between 1 and 10 kpc for X-rays only and at about 50 kpc for the full spectrum. This case also reveals the inherently ``fuzzy" ionization fronts associated with X-rays, relative to the sharp I-fronts of UV radiation - note the gradual transition in spin temperatures for the X-rays-only case spanning nearly two orders of magnitude in scale. In contrast, the case of the  $10^5$ solar-mass starburst with $10^8$ solar-mass QSO/BH (Figure ~\ref{fig:radiobigBH}) reveals that the cases with and without X-rays barely differ in the location and peak values of $T_s$ and $\delta T_b$ (emission in the latter). This arises directly in the strong contribution of X-rays to the overall ionization budget in this scenario.  Ironically, it seems that the greater the X-ray production of a source, the less likely it is have a \textit{distinguishing} X-ray-related signature at 21 cm.

These results reveal one of the key goals of this paper: the difference in the topology of reionization between X-ray and UV ionization scenarios, and their impact on 21 cm predictions. Although X-rays do penetrate deeper into the IGM than do UV photons (leading to the moderate gains in ionization and temperature mentioned earlier), their ``I-front"s trail the UV I-fronts and therefore the UV-associated 21 cm signal. This could therefore ``blur" the signatures of the growth of ionized bubbles around first-light sources, and alter predictions for observing the percolation of reionization (see the semi-numerical simulations of \citealt{warszawski09} on this point). We note that a cosmological scenario in which X-rays alone are generated is not well-motivated physically. Rather, the figures in this section show that the differing scale-dependent ionization from X-rays and UV photons lead directly to  21 cm signals that can be distinguished from each other.

We consider time evolution in two cases for the same source of  a $10^5$ solar-mass starburst with a $10^6$ solar-mass QSO/BH: at  $z = 10$ for times of 1 Myr and 10 Myr after the burst/QSO turn on (Figure ~\ref{fig:radiostd} and Figure ~\ref{fig:radiostd10Myr}), and at $z=20$ for times of 0.1 Myr and 1 Myr (Figure ~\ref{fig:radiostdz20} and Figure ~\ref{fig:radiostdz20-1Myr}).   The main effects of the advancing I-front with time on the 21 cm signal \textit{at a fixed redshift} are the following: a similar advancing of the spin temperature curve's peak, and therefore that of $\delta T_b$, from a few tens of kpc to about 100 kpc, and, a {\it decreased} peak value in $T_s$. This is mostly due to the rapid falloff in the Ly$\alpha$ flux at increasing radii (going as $r^{-2}$), which leads to a decreased coupling between the gas H~I and the source radiation.  The important role of this Ly$\alpha$ photon coupling is manifested also through the slight increase in the positive values (emission signal) of $\delta T_b$ and the increased {\it negative} values of $\delta T_b$ (absorption signal at 21 cm) at $z=20$ relative to $z=10$, arising from the closer location of the I-fronts to the source with increasing redshift. These effects are discussed in more detail in the next section.

\subsection{Comparison with Other Works}

Here, we compare our results and model assumptions with those from papers in the recent literature addressing X-ray and/or helium ionization, and the resulting 21 cm signals. We find that our results are, for the most part, in agreement with the findings of other groups when we make similar model assumptions. We also comment on the theoretical assumption of  passive X-ray production tied to star formation at high redshifts.

\begin{figure}
\begin{center}
\includegraphics[width=3.0in]{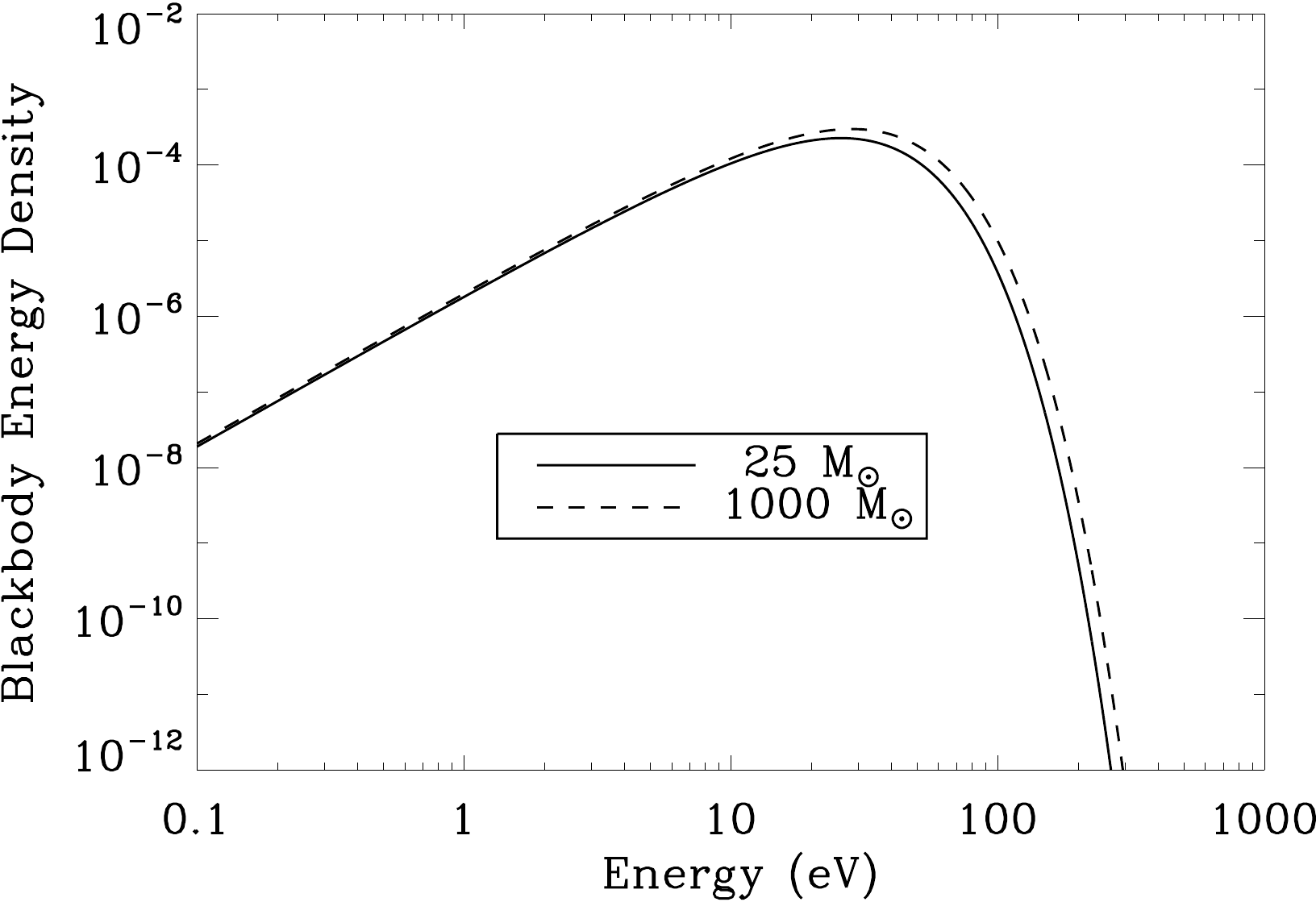}
\caption{A comparison of the blackbody energy output (the Planck energy density, in units of power per unit area per unit solid angle per unit frequency) from a 25 $M_\odot$ and 1000 $M_\odot$ star. Note the relative flatness of the curves at energies of 20--40 eV;  beyond 100 eV the curves decline steeply.}
\label{fig:starsbb}
\end{center}
\end{figure}

In \citet{chen04} and \citet{chen08}, the emergent spectrum is based on a radiative transfer calculation starting with a stellar blackbody spectrum. There is no stated definition to distinguish between X-rays and UV radiation, so that (as in some works on this topic) it is unclear where the X-ray/UV  photon boundary lies. To compare their results with ours, we started with the blackbody spectrum from equation 8 in \citet{chen08}.  The range of Pop~III star masses that they consider (25-800 M$_\odot$) leads to a relatively narrow range of blackbody temperatures, $T_{\rm eff} \sim 1.06 \times 10^5$--$1.17 \times 10^5$ K. In Figure ~\ref{fig:starsbb}, we show the blackbody  energy output (the Planck energy density, in units of power per unit area per unit solid angle per unit frequency) as a function of energy for a 25 $M_\odot$ and a 1000 $M_\odot$ star.  There is little difference between the two cases -- essentially nearly all Pop~III stars have the same energy output \citep{bromm01,tumlinson03}.  

However, what is relevant here for us is the cutoff between UV and X-ray photons. The strict definition of X-rays has a lower limit of 120 eV for X-ray energies. In Figure ~\ref{fig:starsbb}, we see that the energy curves are relatively flat for energies of 20--40 eV and start to decline steeply  above 100 eV.  We were able to reproduce Figure 1 in \citet{chen08} only for the X-ray threshold  energy lying at about 30 eV. Such ``X-rays" can make a substantial addition to the UV-only ionization case, owing to the large numbers of photons below 100 eV. However, placing the cutoff at 100 eV or higher (where the spectrum is down by a factor of $\sim$ 100 relative to the peak), leads to the results in our earlier ionization figures, where X-rays can have a significant (but not dramatic) impact on IGM ionization and temperature.

Also, \cite{chen08} consider a single 200 M$_\odot$ star embedded in a galaxy, versus our treatment of a starburst and/or QSO as a point source in the IGM. The I-fronts in their work are therefore a factor of 10--20 closer to the source than in our results, leading directly to a lower Ly$\alpha$ flux in comparison at large scales in our work. Consequently, the 21 cm {\it absorption} signal induced by the Ly$\alpha$ photons in our calculations is weaker relative to that in \citet{chen08}\footnote{See these authors' discussion in Sec. 2 of their paper of the typical size of Ly$\alpha$ spheres in their work being a few tens of kpc, and of their assumption that the fraction of X-ray energy converted to Ly$\alpha$ photons is 100\%.} or, e.g., \citet{Thomas:08}. This reduced signal is seen as a minor dip, rather than a larger trough, in the 21 cm brightness temperature beyond the I-front location in the right panels of the figures for the $z=10$ cases in Section 3.2. We have checked this by artificially placing the I-fronts in our cases in the 21 cm calculations at closer radii (by $\sim$ a factor of 10) and are able to reproduce the 21 cm brightness temperature absorption signal of \cite{chen08} and other works. Note  that for the $z=20$, $t$ = 0.1 Myr  case  (Figure ~\ref{fig:radiostdz20}), the absorption trough becomes more noticeable as the I-front has not advanced as far. This verifies the critical role of the invere square dropoff of the Ly$\alpha$ photon flux with distance from the source for 21 cm absorption (discussed in an earlier section). Note also that our results for the predicted amplitudes of the spin temperatures and the 21 cm {\it emission} signal are in agreement with other papers in the literature.

In \citet{Bolton:09}, \citet{McQuinn:09}, \cite{furlanetto08a} and \cite{furlanetto08b}, the authors focus on helium reionization by quasars at  $z \sim 3$. \cite{furlanetto08b} do not include the effects of X-ray heating in their calculations of helium ionization. \citet{Bolton:09} find a relatively modest gain in IGM temperature (of order $10^4$ K) resulting from hard radiation, partly owing to the heating in underdense parts of the IGM (particularly fossil He~III regions) achieving their maximal heating early on in the process of reionization (see also \citealt{Venkatesan:03a} on this point). This maximum IGM temperature of $\sim 10^4$ K (comparable to the results of \citealt{McQuinn:09}) lies within the range of our findings, with the caveat that at $z \sim 3$ the IGM is far less dense than at $z \sim$ 10--20,  and, additionally, the IGM hydrogen is completely reionized at $z \sim$ 3, freeing up some of the UV photons and secondary electrons from He~I ionization. We also compared our results with \citet{kuhlen06} - these authors do not include Ly$\alpha$ coupling in their 21 cm calculations but we approximately reproduce their results on spin temperature values.

\citet{Thomas:08} have examined the feedback from early stellar populations and quasars, and the associated 21 cm signature. We found that we were unable to reproduce many of their results, including the high level of X-ray heating ($T > 10^5$--$10^6$ K) as well as  the results in their Figures 12--13.    Some of this may arise from incomplete models of the high-$z$ galaxy distribution and that the stellar spectra have been simplified as a blackbody source. We do however find somewhat similar IGM temperatures and 21 cm brightness temperature values close to those computed in \citet{Ripamonti:08}, although these authors focus on X-rays from BHs in the pre-reionization IGM. A direct comparison is challenging as we do not have a full cosmological calculation in this paper with a halo distribution function characterizing the ionization feedback.

Last, \citet{ricotti05} consider the formation of a strong X-ray background generated at $z \ga 10$ by early black holes, with the specific aim of explaining the  $WMAP$  results of a (at the time high) electron scattering optical depth. In their work, the X-rays are created by accretion onto ``seed" black holes that are assumed to have formed from earlier generations of Pop~III stars. Hence, the total X-ray emissivity is proportional to the total mass in such black holes in high-$z$ halos, which is in turn proportional to the total mass in Pop III stars. That is, the production rate of X-rays is tied effectively to the star formation efficiency in high-$z$ galaxies (through the black hole accretion rate). This is an assumption made in a number of  papers, e.g., \citet{santos08} who examined the role of inhomogeneous X-ray and Ly$\alpha$ radiation fields at $z>$10 for 21 cm signatures of H reionization. However, it remains to be seen how well this series of connections hold at the low black hole masses anticipated in the first galaxies (e.g., do these low-mass  BHs even accrete at the Eddington rate?). The Magorrian relation may not hold at low to moderate BH masses in galaxies \citep{Greene10},  making the scaling of X-ray production with star formation rates and BH masses more ambiguous  at low BH masses.

\citet{ricotti05}, like \citet{chen08}, do not explicitly distinguish between soft X-rays and hard UV photons in their calculations. In these works, the boundary between UV and X-ray photons is related  to the local column density of absorbers and the emergent power spectrum after radiative transfer, with the column density being a free parameter. Another important related parameter is the escape fraction of ionizing radiation, $f_{\rm esc}$, for X-ray and UV photons, which is effectively calculated locally through the emerging flux at each radius (or cell) in our work and for the above papers. Variations in parameters such as the local absorber column density, $f_{\rm esc}$ and reduced gas densities within galaxies owing to feedback effects could  harden the source spectrum within the ionization bubbles, leading to potentially higher temperatures than we have found here. 
Although there is no straightforward way within the scope of our semi-analytic work to directly reproduce local $f_{\rm esc}$ and density-feedback effects from numerical simulations,
we mention these caveats and note that we {\it are} able to reproduce the results of \citet{chen08} by lowering the X-ray/UV boundary to 30 eV or placing the I-fronts closer to the source, both of which effectively harden the local ionizing spectra.  Last,  \citet{ricotti05},  \citet{chen08}, and \citet{Thomas:08} have a fully cosmological calculation that keeps track of the evolving spectra of stellar and QSO populations. Therefore, soft X-rays from high-$z$ sources are redshifted and can become important for hard UV ionization at later epochs. Our current results do not factor in this effect, but we plan to extend this work in the near future to fully cosmological calculations that include realistic galaxy profiles and the redshift evolution of galaxy halos and their radiation fields.

\section{Conclusions}

We have examined the effects of X-rays from high-redshift quasars and
stars when acting in combination with hard UV ionizing radiation from
these sources. We find that, relative to hard UV radiation, X-rays may
not dominate the ionization and thermal history of the IGM, and
contribute modest increases to the IGM ionization at $z \sim$ 10 and
contribute of order $10^3$--$10^5$K to the IGM temperatures. This is
in contrast with some earlier works in which X-rays could cause IGM
heating up to 10$^6$ K and near-total reionization at $z$=10--20.
While some of this may be due to our simplified models, we believe
that most of the difference between our results (where we include the
X-rays coming from individual sources), and those of other works
deriving high IGM temperatures and ionization from X-rays at
$z$ = 10--20, arise from the latter's assumption of strong X-ray production
that is tied to the star formation rate at high redshift.

We also examined the 21 cm signatures of various cases involving combinations of stars and black hole masses, and find that the 21 cm signal of X-ray versus UV ionization could be distinct, resulting from their differing contributions to the topology of reionization. We find that the brightness temperature emission expected from X-rays alone occur at smaller scales than that from UV radiation. The different spatial scales at which they manifest may therefore  ``blur" the 21 cm signature of the percolation of reionization around early halos, depending on whether a cosmic X-ray or UV background built up first. An X-ray background may not  significantly precede a UV background, as a typical X-ray photoionization timescale exceeds the Hubble time for $z \ga$ 10. From our simpified treatment, it is unclear whether there is  a cosmological epoch when the IGM's thermodynamic and ionization properties are determined mostly by X-rays. The role of X-rays versus hard UV radiation can also be tested through their interactions with the CMB, where the relative strengths of their contributions to reionization as well as the redshifts that they dominantly contribute at can be constrained through the CMB polarization power spectrum at large angular scales. The currently operating all-sky CMB mission $Planck$ may be able to distinguish such scenarios. 

For sufficiently hard radiation from sources, the H~II and He~III I-fronts may lie very close to each other. Although our calculation is 1D in nature, this result will impact the escape fraction of ionizing radiation from primordial galaxies, and the geometry of bubbles and chimneys as ionization proceeds from these galaxies. We hope to examine this problem in a future work.

To further explore the evolution of first-light sources and the IGM with redshift, we will extend our current calculations using the Galacticus code  to a fully cosmological framework that includes evolving dark matter halos, galaxy/BH formation, and evolving stellar/QSO populations with time-dependent radiation fields. This will permit a self-consistent calculation of IGM reionization, and allow us to derive predictions for the growth and evolution of a cosmologically representative distribution of ionized bubbles as a function of redshift. We can also calculate the bubbles' thermal properties, as well as the statistical properties of the bubble population, such as the mean size of ionized and neutral regions and power spectra of 21 cm emission or absorption relative to the CMB (utilizing the known correlation properties of the dark matter halos which host the sources).  Such predictions will be tested by data from CMB space telescopes such as {\it Planck}, and ground-based radio telescopes that are designed to map the percolation of reionization around first-light sources. These observations, coupled with our detailed theoretical predictions, will additionally place strong constraints on the populations of ionizing sources at intermediate to high redshifts and, therefore, on the properties of early generations of galaxies and AGN. The resulting improvements in our understanding of these early objects will permit more robust predictions to be made for other observing programs, such as those of the \textit{ James Webb Space Telescope}, which will probe similar galaxy/QSO populations.

\section*{Acknowledgments} AV gratefully acknowledges support from Research Corporation through the Single Investigator Cottrell College Science Award, and from the University of San Francisco Faculty Development Fund. AJB acknowledges the support of the Gordon \& Betty Moore Foundation. We thank the referee for a constructive report, and Massimo Ricotti, Xuelei Chen and Steve Furlanetto for useful input.

\bibliographystyle{mn2e}
\bibliography{paper}

\newpage 

\newpage 

\end{document}